\documentclass[12pt]{article}

\usepackage{epsfig,epstopdf,latexsym,amsfonts,amsmath,amsthm,amssymb,amsbsy,multirow,slashed,wasysym,textcomp,wrapfig,graphicx,psfrag,booktabs,bbm,comment}

\usepackage[toc]{appendix}

\usepackage{color}
\usepackage{datetime}
\usepackage[
      colorlinks=false,
      linkcolor=darkblue,  
      urlcolor=blue,    
      filecolor=blue,     
      citecolor=red,
linktocpage=true,
      pdfstartview=FitV,
      bookmarksopen=true    
      ]{hyperref}

\ifpdf
\fi

\numberwithin{equation}{section}

\def\oneone{\rlap 1\mkern4mu{\rm l}}
\def\coeff#1#2{\relax{\textstyle {#1 \over #2}}\displaystyle}

\def\IC{\mathbb{C}}

\def\IR{\mathbb{R}}
\def\ZZ{\mathbb{Z}}

\def\cB{{\cal B}}

\def\cK{{\cal K}}

\def\cO{{\cal O}}

\def\cQ{{\cal Q}}

\def\Neql#1{{\cal N}\!=\!{#1}}

\definecolor{cardinal}{rgb}{0.6,0,0}
\definecolor{darkgreen}{rgb}{0,0.5,0}
\definecolor{golden}{rgb}{0.92, 0.7, 0}
\definecolor{midnight}{rgb}{0, 0, 0.5}
\definecolor{darkblue}{rgb}{0.2, 0, 0.8}

\begin{document}

\begin{titlepage}

 \begin{flushright}
IPhT-T13/277\\
DCPT-13/53
\end{flushright}

\bigskip
\bigskip
\bigskip

\centerline{\Large \bf On the Oscillation of Species}
\smallskip

\medskip
\bigskip
\bigskip
\centerline{{\bf Iosif Bena$^1$,   Simon F. Ross$^2$ and Nicholas P. Warner$^{1,3,4}$}}
\centerline{{\bf }}
\bigskip
\centerline{$^1$ Institut de Physique Th\'eorique, }
\centerline{CEA Saclay, CNRS-URA 2306, 91191 Gif sur Yvette, France}
\bigskip
\centerline{$^2$ Centre for Particle Theory, Department of Mathematical Sciences }
\centerline{Durham University, South Road, Durham DH1 3LE, UK}
\bigskip
\centerline{$^3$ Department of Physics and Astronomy}
\centerline{University of Southern California} \centerline{Los
Angeles, CA 90089, USA}
\bigskip
\centerline{$^4$ Institut des Hautes Etudes Scientifiques}
\centerline{91440 Bures-sur-Yvette, France}
\bigskip
\bigskip
\centerline{{\rm iosif.bena@cea.fr,~s.f.ross@durham.ac.uk,~warner@usc.edu} }
\bigskip
\bigskip

\begin{abstract}

\noindent We describe a new class of BPS objects called magnetubes: their supersymmetry is determined by their magnetic charges, while their electric charges can oscillate freely between different species. We show how to incorporate these objects  into microstate geometries and create BPS solutions  in which the  charge  densities rotate through different $U(1)$ species as one moves around a circle within the microstate geometry. 
Our solutions have the same ``time-like'' supersymmetry as five-dimensional, three-charge black holes but, in various parts of the solution, the supersymmetry takes the ``null''  form that is normally associated with magnetic charges. It is this property that enables the species oscillation of magnetubes to be compatible with supersymmetry.  We give an example in which the species oscillate  non-trivially around a circle within a microstate geometry, and yet the energy-momentum tensor and metric are completely independent of this circle:  only the amplitude of the oscillation influences the metric.

\end{abstract}

\end{titlepage}

\tableofcontents
\newpage

\section{Introduction}

Microstate geometries play a double role in the study of solutions in supergravity.  First, microstate geometries involve a mechanism that supports time-independent, smooth, solitonic, horizonless solutions within massless supergravity theories and so such solutions can potentially represent  completely new classes of end-states for stars.  The second role is that there are vast classes of such solutions and these classes come with extremely large moduli spaces.  As a result, the fluctuations of microstate geometries can capture a significant portion of the microstate structure of the  black hole that possesses the same  asymptotic structure at infinity.  To understand the extent to which such classical solutions can capture microstate details it is important to understand and classify  the moduli spaces of solutions and find all the ways in which such solutions can fluctuate.  This has been the focus of much recent work and in this  paper we will exhibit a new and broader class of BPS fluctuations that we call {\it species oscillation}.

The key building block of solutions with species oscillations is a new object of string theory, which we will call the {\em magnetube}. The magnetube has M5 and momentum charges (which are magnetic in five dimensions), as well as several electric M2 charge densities that can oscillate between positive and negative values along the M5-P common direction. The original motivation for studying this object, and species oscillation in general, was the idea that it can be used to create microstate geometries for Schwarzschild (electrically neutral) black holes \cite{Mathur:2013nja}.  The specific magnetube considered in  \cite{Mathur:2013nja} involved a KK-monopole with (NS5, D5) charge specified by a vector at each point along a line in the KKM worldvolume.  This charge vector rotates in the (NS5, D5) charge space as one goes around this line so that the resulting configuration has no overall (NS5, D5) charge. At each point on the line the configuration looks like a piece of a BPS supertube, but at different locations the D5 charge of this object can have opposite signs, and so it does not preserve any supersymmetry associated with these charges.

Our first result is to show that the object found in  \cite{Mathur:2013nja}, as well as the class of M5-P magnetubes that we construct in this paper, are secretly supersymmetric. Normally one does not expect this, because the negative and positive M2 (or D5) charge densities at two different locations will attract and will generically want to collapse into each other. However, the story is much richer: because of the presence of extra species one can arrange the M2-brane densities so that, at each point along the direction in which the charge densities vary, the Killing spinors are those of the (magnetic) M5 branes and momentum, and are not affected at all by the presence of (electric) M2 branes. Hence, the physics of the magnetube is the mirror of the physics of normal supertubes \cite{Mateos:2001qs}: for the magnetube the magnetic (M5 and P) charges control the supersymmetry, and the contributions of the fluctuating electric charges to the Killing spinor equations cancel; for normal supertubes the (electric) M2 branes control the supersymmetry and the (magnetic) M5 and momentum charges can fluctuate arbitrarily, as their contribution to the Killing spinor equations cancel each other.

The original spirit of  \cite{Mathur:2013nja} was to try to bend the infinite magnetube with oscillating species (whose solution they constructed) into a round magnetube, which would be neutral and hence give microstate geometries for the Schwarzschild black hole. It is clear that this would break the supersymmetry of the infinite magnetube, as the resulting configuration will have mass and no charge. However there is also an indirect and more useful way to see how supersymmetry is broken: Bending an infinite tube in $\IR^3 \times S^1 $ into a round one in $\IR^4$ can be realized by adding a Taub-NUT center to the solution and bringing this center close to the tube (in the vicinity of the Taub-NUT center the metric is $\IR^4$). For normal supertubes this procedure is supersymmetric  \cite{Elvang:2005sa,Gaiotto:2005xt,Bena:2005ni}, because the Killing spinors of the Taub-NUT space are compatible with the M2-brane Killing spinors of the supertube. However, the magnetube has M5 and P Killing spinors, and those are not compatible with the Taub-NUT ones\footnote{One can see this easily by compactifying to type IIA string theory, where the Taub-NUT space becomes a D6 brane, and the M5 and P become D4 and D0 branes respectively.}, and hence trying to create a neutral configuration results necessarily in the breaking of the supersymmetry preserved by the infinite magnetube solution.

Our second result is to find a way to embed M5-P magnetubes into some of the known supersymmetric solutions and to obtain new classes of BPS solutions with species oscillation. Using magnetubes and species oscillation to create more BPS solutions may appear contrary to the original spirit of \cite{Mathur:2013nja}, which proposed these objects as building blocks for non-extremal microstate geometries. However, this is not so: our result strengthens the evidence for the existence of magnetubes, and proves that there is no obstruction to bending them, and hence to using them in principle for the purpose originally intended in \cite{Mathur:2013nja}. 

Embedding a magnetube into a supersymmetric solution that has M2 Killing spinors may, at first, seem  impossible because the magnetic Killing spinors of the magnetube are not compatible with the electric M2 Killing spinors preserved by the background. Specifically, given a Killing spinor, $\epsilon$, one can define the  vector $K^\mu = \bar \epsilon \gamma^\mu \epsilon$ and this is always a Killing vector that represents the ``time''-translation generated by the anti-commutator of two supersymmetries.  In electric BPS solutions this Killing vector is time-like and in magnetic BPS solutions this Killing vector is null: We therefore use the terminology ``time-like'' or ``null'' supersymmetry  to refer to these two  distinct and apparently incompatible classes.  One can also verify that the supersymmetry projectors associated with the underlying magnetic and electric charges are also incompatible because the projectors do not commute with one another.

On the other hand, when lifted to six dimensions all supersymmetric solutions give rise to null Killing isometries and so, from this perspective, there should be some possibility of interpolation between such supersymmetries.  Indeed, even in five dimensions there is a crucial loop-hole in the structure of the electric BPS supersymmetries:  The four-dimensional spatial bases are allowed to be ambi-polar, which implies that while the Killing vector associated with the supersymmetry is time-like almost everywhere (and particularly at infinity), there are {\it critical surfaces}, where the signature of the base space changes from $(+,+,+,+)$ to $(-,-,-,-)$, and where the Killing vector becomes null.  We will show that the usual time-like supersymmetries of electric charges actually get infinitely boosted and become magnetic or ``null'' supersymmetries on the critical surfaces. It is therefore possible to introduce species oscillation on magnetubes localized right on top of a critical surface, where the Killing spinors of the background become magnetic and are therefore compatible to those of the magnetubes, preserving supersymmetry globally. 

Besides establishing the existence of magnetubes as fundamental building blocks of black hole microstates, the new families of supersymmetric solutions that we construct are interesting in their own right. For example,  in four dimensions the uniqueness theorems are very stringent but in five and six dimensions there are huge families of microstate geometries with the same asymptotic charges. It is thus interesting to explore the range of possibilities  and our new classes of solution will have electric fields that fluctuate and yet the metric either does not fluctuate at all, or the fluctuation response is locally suppressed due to coherent combinations of fluctuating charge densities.  This class of solutions is, in this sense, similar to $Q$-balls \cite{Coleman:1985ki,Friedberg:1986tq}. The latter solutions are supported by time-dependent matter fluctuations arranged so that the energy-momentum tensor and hence the metric are both time-independent.  The solutions we construct here are not time-dependent but the fields fluctuate as a function of some $S^1$-coordinate, $\psi$, and yet the energy-momentum tensor and metric can be arranged to remain invariant along $\psi$.  More generally, one can arrange the fields to fluctuate in such a way that the leading-order perturbations to the energy-momentum tensor near the source remain $\psi$-independent with the fluctuations only becoming visible at sub-leading orders.

We will formulate species oscillation in five-dimensions using a $T^6$ compactification of M-theory that is essentially described in \cite{Giusto:2012gt}.  There will be one spectator M2 brane wrapping a fixed $T^2$ and the species oscillation will take place between four classes of M2-branes that wrap the remaining $T^4$ in different ways.  One can also reduce this to IIA by compactifying on an $S^1$ inside the $T^2$ of the spectator M2, converting it to an F1 string while the oscillating M2's become oscillating D2's.  If one T-dualizes on the  circle wrapped by the F1, one obtains a IIB compactification in which the F1 has become momentum charge and the oscillating species are simply three sets of D3-branes that intersect along a common circle while wrapping the $T^4$ in exactly the same manner as the original M2's.  The importance of the IIB frame is that it is only in this frame that our solutions will be completely smooth microstate geometries.  Our formulation also has the advantage of having all the oscillating species originating from the same type of branes. It is also possible to relate some of our magnetube solutions to the smooth IIB frame magnetubes obtained in \cite{Mathur:2013nja}, by a duality sequence that we will discuss in Section \ref{Sect:Ten}.

In Section \ref{Sect:SusyTypes} we discuss the two classes, time-like and null, of supersymmetry and their relation to electric  and  magnetic BPS solutions in M-theory.  We also show how both types of supersymmetry can be present in BPS solutions with ambi-polar base metrics.  In Section \ref{Sect:AddVMs} we discuss the relevant five-dimensional $\Neql{2}$ supergravity theories and their origins in M-theory.  In Section \ref{Sect:SixDim} we discuss the six-dimensional uplifts and their relation to IIB supergravity.  We also discuss the regularity conditions that make supertubes into smooth microstate geometries in six dimensions.  Section \ref{Sect:Temp} contains a ``template'' microstate geometry which is a standard, non-oscillating microstate geometry with two Gibbons-Hawking (GH) geometric charges and a ``supertube'' on the critical surface between the GH charges. In reality this ``supertube'' is actually a magnetube because it preserves a null supersymmetry. Indeed, the constructions of supertubes and magnetubes are mathematically parallel and, following the logic of Section  \ref{Sect:SusyTypes}, one can think of magnetubes as infinitely-boosted supertubes. In Section \ref{Sect:Temp} we also  check the regularity and asymptotic structure of this magnetube solution.  In Section \ref{Sect:SOsoln} we introduce species oscillation into the template magnetube and give an example in which the metric does not oscillate.  We also see that the metrics of the oscillating solutions are almost identical to those of the template solutions and are therefore regular. In Section \ref{Sect:Modes} we give details of the Green's functions and explicit mode functions that are the essential, though technical, part of our solutions with species oscillation.  While our examples involve only very simple microstate geometries, they will nevertheless provide a good local model of what we expect from species oscillation in a generic bubbling solution. Finally we make some concluding remarks in Section \ref{Sect:Conclusions}.

\section{Merging the two types of supersymmetry}
\label{Sect:SusyTypes}

In this section we illustrate one of the key ingredients of species oscillation:  The fact that one can mix both time-like and null supersymmetric components within a single, five-dimensional solution.  We will assume some familiarity with the five-dimensional microstate geometries that have been constructed over the last few years.  One can find discussions of this in, for example,  \cite{Bena:2005va,Berglund:2005vb,Bena:2007kg,Gibbons:2013tqa}.  In Section \ref{Sect:AddVMs} we will discuss the general form of  five-dimensional $\Neql{2}$ supergravity  coupled to an arbitrary number of vector multiplets and this can also serve as something of a review of the relevant supergravity structure but in  greater generality than is needed to understand the merging of the two types of supersymmetry.  Since the latter idea is something rather new, we have chosen to show how it works here first before diving more deeply into the technicalities of the more complicated supergravity theories that are needed to implement species oscillation.  For newcomers to the microstate geometry program it might be useful to review Section \ref{Sect:AddVMs} first.

\subsection{Time-like and null supersymmetry of M2 and M5 branes}

In both eleven-dimensional and five-dimensional supergravity there are two distinct classes of supersymmetry, which are classified by whether the associated ``time'' translation invariance is actually time-like or null.  More precisely, if $\epsilon$ is the residual supersymmetry then the vector
\begin{equation}
\zeta^a ~\equiv~ \bar \epsilon \Gamma^a \epsilon \,, 
\label{KVform}
\end{equation}
is necessarily a Killing vector, and it can either be time-like or null \cite{Gauntlett:2002nw,Gauntlett:2002fz}.  It is relatively easy to see that the vector is dominated by its time component because, in frame indices, one has $\zeta^0 =  \epsilon^\dagger \epsilon$.  To investigate the other components one needs to know a little more about the structure of the supersymmetry.

One the simplest descriptions of  BPS three-charge  black holes in five dimensions is obtained by compactifying eleven-dimensional supergravity on a six-torus, $T^6$.  The charges are then carried by three sets of mutually BPS M2-branes and thus the supersymmetries obey the projection conditions 
\begin{equation}
\Gamma^{056} \, \epsilon ~=~  \Gamma^{078}\,  \epsilon ~=~ \Gamma^{09\, 10} \, \epsilon ~=~ \epsilon \,, 
\label{proj1}
\end{equation}
where the five-dimensional space-time is coordinatized by $(t=x^0,x^1, x^2, x^3,x^4)$ and the $T^6$ is coordinatized  by $(x^5,\dots,x^{10})$.   The  microstate geometries corresponding to such black holes are, by definition, required to have the same supersymmetries, in that they also satisfy  (\ref{proj1}). 
One should also note that because $\Gamma^{01\dots 9\,10} = \oneone$,  the conditions (\ref{proj1}) also imply
\begin{equation}
\Gamma^{1234} \, \epsilon ~=~ \epsilon \,,
\label{proj1a}
\end{equation}
which constrains the holonomy of the spatial base.

One can now insert the  $\Gamma^{0cd}$ into (\ref{KVform}) and commute through the $\Gamma^a$. The fact  that $\Gamma^a$, for $a\ne 0$, anti-commutes with at least one of the $\Gamma^{0cd}$ means that  $\zeta^a=0$ for $a\ne 0$ and so the Killing vector for such black holes and their microstate geometries is necessarily time-like. The BPS three-charge black holes and their microstate geometries are dominated by their electric charge structure but can carry non-zero, ``dipolar'' magnetic charge distributions whose fields fall off too fast to give any net charge at infinity. 

There is a simple magnetic dual of this picture in which the M2 branes are replaced by M5 branes. We will take the M5 branes to have a common direction, $\psi = x^4$, which, for the present, will be flat and either infinite or a trivially fibered $S^1$.  The M5-brane supersymmetries obey the projection conditions 
\begin{equation}
\Gamma^{0\psi5678} \, \epsilon ~=~  \Gamma^{0\psi569\,10}\,  \epsilon ~=~ \Gamma^{0\psi789\,10} \, \epsilon ~=~ \epsilon \,, 
\label{proj2}
\end{equation}
and these also  imply the $\psi$-momentum projection condition:
\begin{equation}
\Gamma^{0\psi} \, \epsilon ~=~  \epsilon \,.
\label{proj2a}
\end{equation}
Again one can  insert the  $\Gamma^{0\psi}$ into (\ref{KVform}) and commute through the $\Gamma^a$ and conclude that $\zeta^a=0$ for $a\ne 0,\psi$. Moreover  (\ref{proj2a}) implies $\zeta^0=\zeta^\psi$ and so the Killing vector is necessarily null.  Thus these systems of M2 branes and M5 branes are exemplars of the two classes of supersymmetric systems in five and eleven dimensions.

For our purposes, it is important to note that, just as the  BPS three-charge black holes and  microstate geometries can be given non-zero, ``dipolar'' magnetic charge distributions whose fields fall off too fast to give any net charge at infinity, the M5 brane system can be given non-zero, ``dipolar'' {\it electric} charge distributions whose fields fall off too fast to give any net charge at infinity.  In this way one can make BPS magnetubes that have no net electric charge and this observation will lie at the heart of supersymmetric species oscillation.

\subsection{The ``classic'' solutions with  time-like supersymmetry}
\label{Sect:classic}

So far we have considered rather simple classes of BPS configurations and, as we will see, more complex BPS backgrounds can lead to mixtures of both types of supersymmetry.  Specifically we want to consider backgrounds that have become the standard fare for the construction of microstate geometries\footnote{For later convenience we are going to label the species of M2 branes and the corresponding fields by $0,1,2$ and not use the more usual labeling,  $1,2,3$.  We have therefore taken the standard description and replaced the label $3$ by $0$ everywhere.}.  The eleven-dimensional metric  has the form:
\begin{align}
 ds_{11}^2  ~=~  - ( Z_0 Z_1  Z_2 )^{-{2\over 3}}  (dt+k)^2  & + ( Z_0 Z_1  Z_2)^{1\over 3} \, ds_4^2   +     (Z_0 Z_2 Z_1^{-2})^{1\over 3}  (dx_5^2+dx_6^2) \nonumber \\
 & +  ( Z_0 Z_1  Z_2^{-2} )^{1\over 3} (dx_7^2+dx_8^2)    +   (Z_1 Z_2  Z_0^{-2} )^{1\over 3} (dx_9^2+dx_{10}^2) \,,
\label{elevenmetric}
\end{align}
where the four-dimensional space-time metric has the standard Gibbons-Hawking (GH)  form:
\begin{equation}
ds_4^2 ~=~  V^{-1} \, (d\psi + A)^2  ~+~  V\, d\vec y \cdot d\vec y   \,,
\label{GHmetric}
\end{equation}
with 
\begin{equation}
\vec \nabla \times \vec A ~=~ \vec \nabla V   \,.
\label{AVreln}
\end{equation}

The eleven-dimensional Maxwell three-form potential is given by
\begin{equation}
C^{(3)}  = A^{(1)} \wedge dx^5 \wedge dx^6 ~+~  A^{(2)}   \wedge
dx^7 \wedge dx^8 ~+~ A^{(0)}  \wedge dx^9 \wedge dx^{10}  \,,
\label{Cfield}
\end{equation}
with the five-dimensional Maxwell fields, $A^{(I)}$, given by: 
\begin{equation}
A^{(I)} ~=~ -Z_I^{-1} \, (dt+k) ~+~   \Big( \frac{K^I}{V}\Big) \, (d\psi + A) ~+~ \vec \xi^{(I)} \cdot d\vec y \,,
\label{AIfield}
\end{equation}
with  $\vec \nabla \times \vec \xi^{(I)} ~=~ - \vec \nabla K^I$.

The whole solution is then specified by eight harmonic functions, $V$, $K^I$, $L_I$ and $M$, on $\IR^3$, and these are typically taken to have the form:
\begin{equation}
 V = \varepsilon_0 ~+~ \sum_{j=1}^N \,  {q_j  \over r_j} \,, \qquad K^I ~=~ k^I_0 ~+~  \sum_{j=1}^N \, {k_j^I \over r_j} \,,
\label{VKdefn}
\end{equation}
\begin{equation}
 L_I ~=~ \ell^I_0 ~+~  \sum_{j=1}^N \, {\ell_j^I \over r_j} \,, \qquad
M ~=~ m_0 ~+~  \sum_{j=1}^N \, {m_j \over r_j} \,,
\label{LMdefn}
\end{equation}
where $r_j \equiv  |\vec y  - \vec y^{(j)} |$.

The electrostatic potentials and warp factor functions, $Z_I$, are given by 
\begin{equation}
Z_I ~=~ \coeff{1}{2}  \, C_{IJK} \, V^{-1}\,K^J K^K  ~+~ L_I \,,
\label{ZIform}
\end{equation}
where  $C_{IJK} ~\equiv~ |\epsilon_{IJK}|$.
The angular-momentum vector, $k$, decomposes into: 
\begin{equation}
k ~=~ \mu\, ( d\psi + A   ) ~+~ \omega \,, 
\label{kansatz}
\end{equation}
where:
\begin{equation}
\mu ~=~ \frac{1}{6\, V^2} \, C_{IJK}\,  K^I K^J K^K ~+~
{1 \over 2 \,V} \, K^I L_I ~+~  M\,. 
\label{muform}
\end{equation}
and where $\omega$ is the solution to:
\begin{equation}
\vec \nabla \times \vec \omega ~=~  V \vec \nabla M ~-~
M \vec \nabla V ~+~   \coeff{1}{2}\, (K^I  \vec\nabla L_I - L_I \vec
\nabla K^I )\,.
\label{omegeqn}
\end{equation}
%

\subsection{Merging the two classes of supersymmetry}
\label{Sect:Merge}

One can uplift the five-dimensional supergravity solutions to six dimensions by using one of the vector fields to make a KK fiber.  We will discuss this extensively in section \ref{Sect:Temp} but here we simply want to note that in the six-dimensional uplift there is only one kind of supersymmetry \cite{Gutowski:2003rg,Cariglia:2004kk}: the null supersymmetry with $\zeta^\mu \zeta_\mu  = 0$.  The two types of supersymmetry in five dimensions must therefore emerge from the details of the compactification to five dimensions.  In this sense one might naturally expect to unify the two classes of supersymmetry,  and we will now elucidate how this can be seen directly from the five-dimensional perspective. 

As have been extensively noted elsewhere  \cite{Giusto:2004kj,Bena:2005va,Berglund:2005vb,Bena:2007kg,Gibbons:2013tqa}  the metric on the four-dimensional base is allowed to be ambi-polar, that is, it is allowed to change signature from $(+,+,+,+)$ to $(-,-,-,-)$. In spite of this, the five-dimensional and eleven-dimensional metrics are smooth and Lorentzian.  In particular, this means that $V$ is allowed to change sign and the surfaces where $V=0$ are called {\it critical surfaces}.  One can make a careful examination of the metric  (\ref{elevenmetric}) and Maxwell fields (\ref{AIfield}) and show that the apparently singular terms involving negative powers of $V$ actually cancel out on critical surfaces, leaving a smooth background.  It is also important to note that the functions $Z_I V$ must satisfy
\begin{equation}
Z_I \, V ~>~  0 
\label{ZIVpos}
\end{equation}
everywhere in order for the metric (\ref{elevenmetric}) to be real and Lorentizian.

This is not to say that nothing is happening at the critical surfaces.  Observe the the norm of the Killing vector, $\zeta = \frac{\partial}{\partial t}$, is simply 
\begin{equation}
\zeta^\mu \,  \zeta_\mu   ~=~    - ( Z_0 Z_1  Z_2 )^{-{2\over 3}} ~=~    - ( (Z_0 V)  (Z_1 V) (Z_2 V) )^{-{2\over 3}} \, V^2 \,.
\label{Ksquared}
\end{equation}
Since the $Z_I V$ are globally positive, this means that $K^\mu$ is time-like except on critical surfaces, where it becomes null.  Thus the supersymmetries are time-like almost everywhere, except on critical surfaces where they momentarily become null.  

It is therefore possible, in principle, to have both types of supersymmetry within one class of solutions.  Indeed, black-hole microstate geometries have the same time-like supersymmetries at infinity as a black hole and generically have critical surfaces in the interior of the solution where the supersymmetries become null. It is instructive to see how this comes about in detail.

Introduce the obvious frames for the five-dimensional metric:
\begin{equation}
e^0  ~\equiv~    Z^{-1} \, (dt+k) \,, \qquad e^1  ~\equiv~ (ZV)^{\frac{1}{2}} \, V^{-1} \,  (d\psi + A) \,, \qquad e^{a+1} ~\equiv~ (ZV)^{\frac{1}{2}} \,   dy^a \,,
\label{obvframes}
\end{equation}
where 
\begin{equation}
Z  ~\equiv~  V^{-1} \,( (Z_0 V)  (Z_1V)    (Z_2 V) )^{{1\over 3}}  \,.
\label{Zdefn}
\end{equation}
Note that we have chosen to write these quantities in terms of the positive functions, $Z_I V$, so that the fractional powers are unambiguous and have no branch cuts.  Also observe that the  frames $e^0$ and $e^1$ are, respectively, degenerate or singular  on  critical surfaces.

Define null frames:
\begin{align}
e^+  ~\equiv~ &  V\, (e^0 + e^1) \,, \qquad \qquad e^-  ~\equiv~   V^{-1} \,(- e^0 + e^1) \\
\Leftrightarrow \quad  e^0  ~\equiv~ &  \coeff{1}{2} \,(V^{-1} \, e^+ - V\,  e^-) \,, \qquad e^1 ~\equiv~    \coeff{1}{2} \,(V^{-1} \, e^+ + V\,  e^-)  \,.
\label{epmdefn}
\end{align}
and thus 
\begin{equation}
ds_5^2  ~=~  -(e^0)^2 ~+~ \sum_{j=1}^4 \, (e^j)^2 ~=~ e^+ \, e^- ~+~  \sum_{j=a}^3 \, (e^{a+1})^2 \,.
\label{metfram}
\end{equation}

It is also convenient to introduce the quantity
\begin{equation}
\cQ  ~\equiv~  Z_0 Z_1 Z_2 V ~-~ \mu^2 \, V^2 \label{Qdefn} 
\end{equation}
and recall  \cite{Bena:2005va,Berglund:2005vb,Bena:2007kg}  that the general expression for $\cQ$ can be written in terms of the $E_{7}$ invariant that determines the four-dimensional horizon area for black-hole solutions:
\begin{eqnarray} \cQ   &=& - M^2\,V^2   ~-~ \coeff{1}{3}\,M\,C_{IJK}{K^I}\,{K^J}\,{K^K}
~-~ M\,V\,{K^I}\,{L_I} ~-~ \coeff{1}{4}(K^I L_I)^2 \nonumber \\
&&~+~ \coeff{1}{6} \,V C^{IJK}L_I L_J L_K ~+~\coeff{1}{4}\, C^{IJK}C_{IMN}L_J L_K K^M K^N \,. \label{Qreduced}
\end{eqnarray}
In particular, it follows from  (\ref{Qreduced}) that $\cQ$ is smooth across the  critical ($V=0$) surfaces.

The whole point is that while the frames $e^0$ and $e^1$ are singular on the critical surfaces, $e^+$ and $e^-$ are {\it smooth} on these surfaces.  To see this, note that 
\begin{align}
e^+ ~=~ & (ZV)^{-1} V^2 (dt + \omega)   ~+~ \big( \, (ZV)^{-1} \, (\mu \, V^2) + (ZV)^{\frac{1}{2}}  \, \big) \, (d\psi + A) \,, \label{eplussimp} \\
e^- ~=~ & -(ZV)^{-1} (dt + \omega)   ~+~ (ZV)^{-1} \, \, \big( \,V^{-2}  \, (ZV)^{\frac{3}{2}} - \mu \, \big) \, (d\psi + A) \,. \label{eminussimp} 
\end{align}
It follows from (\ref{ZIform}) and (\ref{muform})  that $ZV$ and $\mu V^2$ are well behaved on critical surfaces and thus $e^+$ is manifestly well behaved. Now observe that 
\begin{equation}
\cQ  ~=~  \big( \,V^{-2}  \, (ZV)^{\frac{3}{2}} - \mu \, \big) \,\big( \,(ZV)^{\frac{3}{2}} + \mu \,V^2 \, \big) \, \label{Qfac} 
\end{equation}
and that as $V\to 0$, the second factor is finite. Since $\cQ$ is finite as $V\to 0$, it follows that the first factor in  (\ref{Qfac}) must be similarly finite and hence $e^-$ is finite on critical surfaces.

More explicitly, define
\begin{equation}
\cK ~=~ (K^0 K^1 K^2)^{\frac{1}{3}} \, \label{cKdefn} 
\end{equation}
and note that global positivity of $Z_IV$ implies that $K^0 K^1 K^2 >0$ on critical surfaces and so $\cK$ is real and smooth across these surfaces.
As $V\to 0$, one finds
\begin{equation}
e^+ ~\to~  2  \,\cK \,  (d\psi + A) \,,  \qquad  e^- ~\to~  -\cK^{-2} (dt + \omega)   ~+~ \coeff{1}{2} \, \cK^{-5} \, \cQ \, (d\psi + A) \,, \label{epmlim} 
\end{equation}
which are clearly regular as $V \to 0$.

Finally, define new canonical Lorentzian frames via
\begin{align}
\hat e^0  ~\equiv~ & \coeff{1}{2} \, (e^+ - e^-) ~=~  \coeff{1}{2} \, (V +V^{-1} )\, e^0 +\coeff{1}{2} \, (V -V^{-1} )\, e^1 \,, \\
\hat e^1  ~\equiv~  & \coeff{1}{2} \, (e^+ + e^-) ~=~  \coeff{1}{2} \, (V - V^{-1} )\, e^0 +\coeff{1}{2} \, (V +V^{-1} )\, e^1   \,.
\label{ehatdefn}
\end{align}
These are also regular as one crosses the $V=0$ surface but more importantly, they manifestly represent a Lorentz boost of the original frames (\ref{obvframes}) with boost parameter, $\chi$, determined by
\begin{equation}
\cosh \chi  ~=~   \coeff{1}{2} \, \big(|V| +|V|^{-1}\big)  \,, \qquad \sinh \chi  ~=~    \coeff{1}{2} \, \big( |V| - |V|^{-1}\big)  \,. \label{boost1} 
\end{equation}
Thus one can see that the original frames are singular at $V=0$ precisely because they are infinitely boosted relative to a smooth set of frames.  This is consistent with the fact that the Killing vector, $\zeta^\mu$, becomes momentarily null at $V=0$.

Now recall the identity
\begin{equation}
\epsilon^\dagger \,  \epsilon~=~ \bar \epsilon \Gamma^0 \epsilon ~=~  \zeta^0 ~=~ Z^{-1} ~=~ (ZV)^{-1} \, V\,,
\label{mageps}
\end{equation}
where we are using frame indices based on (\ref{obvframes}).   This implies that in this singular frame basis the magnitude of $\epsilon$ vanishes as $\cO(|V|^{1/2})$ when $V \to 0$.  Passing to the non-singular frames (\ref{ehatdefn}) requires a Lorentz boost that acts on $\epsilon$:
\begin{equation}
\epsilon  ~\to~   \hat \epsilon ~=~  \exp(\coeff{1}{2} \chi \Gamma^{01}) \,   \epsilon ~=~   \coeff{1}{2}\, |V|^{1/2} \, \big(\oneone + \Gamma^{01} \big) \,   \epsilon~+~  \coeff{1}{2}\, |V|^{-1/2} \, \big(\oneone -  \Gamma^{01} \big) \,   \epsilon\,. \label{boost2} 
\end{equation}
This means that on the $V=0$ surface only the components satisfying 
\begin{equation}
\Gamma^{01}  \,   \epsilon  ~=~ -\epsilon \,,
\label{boostedproj}
\end{equation}
remain finite.  Thus the Killing spinor is time-like everywhere except where $V=0$, where it becomes momentarily null.  

We therefore conclude that if $V\ne 0$ then the supersymmetry is precisely that of the three-charge M2-brane system, but on the $V=0$ surfaces this supersymmetry becomes compatible with the supersymmetry of M5 branes wrapped on the $\psi$-circle and with momentum charge on that circle.  

\subsection{A degenerate limit and a class of null BPS solutions }
\label{Sect:Vzero}

The fact that, despite appearances, the BPS solutions described in Section \ref{Sect:classic} are regular across $V=0$ surfaces enables one to take this somewhat further and find classes of BPS solutions by taking $V\equiv 0$ everywhere.  As one might anticipate from the discussion above, these  solutions have null supersymmetry and are sourced primarily by M5 branes and momentum.

To take this limit one can rescale $V \to \lambda V$ everywhere in the solutions of Section \ref{Sect:classic} and take $\lambda$ to zero.
The metric (\ref{elevenmetric}) simplifies to
\begin{align}
 ds_{11}^2  ~=~ &  - 2\, \cK^{-1} \,d\psi \, \Big(dt +\omega - \coeff{1}{2} \,\cK^{-3} \,\widehat \cQ \, d\psi \Big) ~+~  \cK^{2} \,   d\vec y \cdot d\vec y  +    \Big (\frac{(K^1)^2 }{K^0 \, K^2}\Big)^{1\over 3}  (dx_5^2+dx_6^2) \nonumber \\
 & +   \Big (\frac{(K^2)^2 }{K^0 \, K^1}\Big)^{1\over 3} (dx_7^2+dx_8^2)    +   \Big (\frac{(K^0)^2 }{K^1 \, K^2}\Big)^{1\over 3}  (dx_9^2+dx_{10}^2) \,,
\label{red11metric}
\end{align}
where $\cK$ is defined in (\ref{cKdefn}) and $\widehat \cQ$ is simply $\cQ$ with $V\equiv 0$:
\begin{equation}  
\widehat \cQ   ~=~   -2\,\cK^3 \, M  ~-~ \coeff{1}{4}(K^I L_I)^2  ~+~ \coeff{1}{4}\, C^{IJK}C_{IMN}L_J L_K K^M K^N \,. \label{hatQ}
\end{equation}
The angular momentum vector, $ \omega$, is now determined by the simpler equation
\begin{equation}
\vec \nabla \times \vec \omega ~=~    \coeff{1}{2}\, (K^I  \vec\nabla L_I - L_I \vec \nabla K^I )\,.
\label{hatomega}
\end{equation}
The electromagnetic fields (\ref{AIfield}) reduce to the purely magnetic forms:
\begin{align}
A^{(0)} ~=~  & \frac{1}{K^1 \,K^2}\, \big( K^0 \, L_0 - K^1 \, L_1 - K^2 \, L_2\big)  \,d \psi ~+~   \vec \xi^{(0)} \cdot d\vec y  \,, \nonumber \\ 
A^{(1)} ~=~ & \frac{1}{K^2 \,K^3}\, \big( K^1 \, L_1 - K^2 \, L_2 - K^0 \, L_0\big)  \,d \psi ~+~   \vec \xi^{(1)} \cdot d\vec y \,, \nonumber \\ 
A^{(2)} ~=~  & \frac{1}{K^1 \,K^3}\, \big( K^2 \, L_2 - K^1 \, L_1 - K^0 \, L_0\big)  \,d \psi ~+~   \vec \xi^{(2)} \cdot d\vec y \,.
\label{AIfieldred}
\end{align}
This is the solution that corresponds to the infinite magnetube and, as expected, the Killing vector, $\zeta = \frac{\partial}{\partial t}$, is manifestly null everywhere in the metric (\ref{red11metric}).

\section{Supergravity coupled to four vector multiplets}
\label{Sect:AddVMs}

\subsection{The supergravity action}
Species oscillation requires the addition of extra vector multiplets to the $\Neql2$ supergravity theory employed in Section \ref{Sect:SusyTypes} and we will therefore summarize the relevant aspects of these theories.  Our conventions and normalizations will be those of \cite{Gutowski:2004yv,Gauntlett:2004qy}. 

The action of $\Neql2$, five-dimensional supergravity coupled to $N$ $U(1)$ gauge fields is given by:
\begin{eqnarray}
  S = \frac {1}{ 2 \kappa_{5}} \int\!\sqrt{-g}\,d^5x \Big( R  -\coeff{1}{2} Q_{IJ} F_{\mu \nu}^I   F^{J \mu \nu} - Q_{IJ} \partial_\mu X^I  \partial^\mu X^J -\coeff {1}{24} C_{IJK} F^I_{ \mu \nu} F^J_{\rho\sigma} A^K_{\lambda} \bar\epsilon^{\mu\nu\rho\sigma\lambda}\Big) \,,
  \label{5daction}
\end{eqnarray}
with $I, J =0, \dots, N$.  The extra photon lies in the gravity multiplet and so there are only $N$  independent scalars. It is, however, convenient to parametrize them by $N+1$ scalars $X^I$,  satisfying  the  constraint
\begin{equation}
\coeff{1}{6} \, C_{IJK} X^I \,  X^J \, X^K  ~=~  1 \,.
\label{Xconstr}
\end{equation}
Following standard conventions, introduce
\begin{equation}
X_I ~\equiv~  \coeff{1}{6} \, C_{IJK}   X^J \, X^K \,.
\label{Xdown}
\end{equation}
The  scalar kinetic term can then be written as
\begin{equation}
  Q_{IJ} ~=~  \coeff{9}{2} \, X_I \, X_J ~-~ \coeff{1}{2} \, C_{IJK} X^K \,.
\label{scalarkinterm}
\end{equation}
The Chern-Simons structure constants are required to satisfy the constraint
\begin{equation}
 C_{IJK} \,   C_{J' (LM} \,  C_{PQ)K'}  \, \delta^{J J'} \,  \delta^{K K'} ~=~  \coeff{4}{3} \, \delta_{I(L}\, C_{MPQ)}    \,.
\label{Cconstr}
\end{equation}
It is also convenient to define
\begin{equation}
 C^{IJK}~=~   \delta^{I I'} \,  \delta^{J J'} \,  \delta^{K K'}  \, C_{I'J'K'}    \,.
\label{Cupper}
\end{equation}
Using the constraint (\ref{Cconstr}), one can  show that the inverse, $Q^{IJ}$, of $Q_{IJ}$ is given by:
\begin{equation}
Q^{IJ} ~=~ 2 \, X^I \, X^J ~-~ 6\, C^{IJK} X_K    \,, 
\label{Qinv}
\end{equation}
and one can show that
\begin{equation}
\coeff{1}{6} \, C^{IJK} X_I \,  X_J \, X_K  ~=~  \frac{1}{27} \,.   
\label{Cconstr2}
\end{equation}

We are simply going to follow \cite{Giusto:2012gt,Vasilakis:2012zg} and consider eleven-dimensional supergravity reduced on a $T^6$.  The Maxwell fields descend from the tensor gauge field, $C^{(3)}$, via harmonic $2$-forms on $T^6$.  The structure constants $C_{IJK}$ are then simply given by the intersection form of the dual homology cycles and the $X^I$ are moduli of the $T^6$.

\subsection{The supersymmetry conditions}

We start with the most general stationary five-dimensional metric:
\begin{equation}
ds_5^2 ~=~ -Z^{-2} \,(dt + k)^2 ~+~ Z \, ds_4^2  \,,
\label{metAnsatz}
\end{equation}
where $Z$ is simply a convenient warp factor.   Supersymmetry implies, via the condition (\ref{proj1a}), that the metric $ds_4^2$ on the spatial base manifold, $\cB$, must be hyper-K\"ahler.

One now defines $N+1$ independent functions, $Z_I$ by
\begin{equation}
Z_I ~=~  3 \, Z \, X_I  \,,
\label{ZIdefn}
\end{equation}
and then (\ref{Cconstr2}) implies
\begin{equation}
Z   ~=~  \big(\coeff{1}{6} \, C^{IJK} Z_I \, Z_J \, Z_K\big)^{\frac{1}{3}}\,.   
\label{Zreln}
\end{equation}
It is more convenient to think of the solution as parametrized by the $N+1$ independent scalars, $Z_I$, and that the warp factor is determined by (\ref{Zreln}).

Supersymmetry requires that the Maxwell potentials all have the form
\begin{equation}
A^{(I)}   ~=~  - \coeff{1}{2} \, Z^{-3}\, C^{IJK} \, Z_J \, Z_K (dt +k)  ~+~ B^{(I)}  \,,
\label{AAnsatz}
\end{equation}
where $B^{(I)}$ are purely magnetic components on the spatial base manifold, $\cB$.  One defines the magnetic field strengths accordingly: 
\begin{equation}
\Theta^{(I)}   ~=~ d \, B^{(I)}  \,.
\label{Thetadefn}
\end{equation}

Having made all these definitions, the BPS equations take on their canonical linear form \cite{Bena:2004de}:
\begin{eqnarray}
 \Theta^{(I)}  &=&  \star_4 \, \Theta^{(I)} \label{BPSeqn:1} \,, \\
\nabla^2_{(4)}  Z_I &=&  \coeff{1}{2} \,C_{IJK} \, \star_4 \Theta^{(J)} \wedge
\Theta^{(K)} \label{BPSeqn:2} \,, \\
 dk ~+~  \star_4 dk &=&  Z_I \,  \Theta^{(I)}\,,
\label{BPSeqn:3}
\end{eqnarray}
where $\star_4$ is the Hodge dual in the four-dimensional base metric $ds_4^2$, and  $\nabla^2_{(4)}$ is the (four-dimensional) Laplacian of this metric.

We are, once again, going to take the metric on $\cB$ to be a GH metric (\ref{GHmetric}) and so we will also decompose the vector $k$ according to  (\ref{kansatz}).  The GH metric and the magnetic fluxes will also be as before and thus we will take $V$ and $K^I$ to have the form (\ref{VKdefn}).   However, we will allow the rest of the solution ($Z_I$, $\mu$ and $\vec \omega$) to depend upon all four variables, $(\psi,\vec y)$.   The $Z_I$'s will still have the form 
(\ref{ZIform}) but now the $L_I$ are general harmonic functions on the base $\cB$: 
\begin{equation}
\nabla^2_{(4)} L_I ~=~ 0   \,.
\label{Lharm}
\end{equation}
The BPS solutions in these circumstances have been discussed in \cite{Bena:2010gg,Niehoff:2013kia}.  The last BPS equation, (\ref{BPSeqn:3}), can be written as 
\begin{equation}
( \mu \vec {\mathcal{D}} V - V\vec {\mathcal{D}}  \mu  ) ~+~      \vec {\mathcal{D}}  \times \vec \omega ~+~    V \partial_\psi  \vec \omega  ~=~ -  V\, \sum_{I=1}^3 \, Z_I \, \vec \nabla \big(V^{-1} K^I  \big) \,,
\label{simpcovkeqn}
\end{equation}
where 
\begin{equation}
\vec {\mathcal{D}} ~\equiv~ \vec \nabla ~-~   \vec A \,\partial_\psi   \,. 
\label{simpcovD}
\end{equation}
This BPS equation, (\ref{BPSeqn:3}), has a gauge invariance:  $k \to k + df$ which translates to
\begin{equation}
\mu \to \mu ~+~ \partial_\psi f \,, \qquad  \vec \omega \to  \vec \omega ~+~ \vec {\mathcal{D}} f \,. 
\label{fgaugetrf}
\end{equation}
It is simplest to use a Lorentz gauge-fixing condition, $d\star_4k =0$, which reduces to
\begin{equation}
V^2 \, \partial_\psi \mu   ~+~\vec {\mathcal{D}}  \cdot \vec \omega  ~=~ 0 \,.
\label{Lorgauge}
\end{equation}

The four-dimensional Laplacian can be written as
\begin{equation}
\nabla^2_{(4)}  F ~=~ V^{-1} \big[ V^2 \, \partial_\psi^2  F   ~+~ \vec {\mathcal{D}}  \cdot  \vec {\mathcal{D}}    F \big] \,.
\label{Lapl}
\end{equation}
If one takes the covariant divergence of (\ref{simpcovkeqn}) (using $\vec {\mathcal{D}}$) and uses the Lorentz gauge choice,  one obtains
\begin{equation}
V^{^2}\, \nabla^2_{(4)}    \mu  ~=~   \vec {\mathcal{D}} \cdot \Big( V\, \sum_{I=1}^3 \, Z_I \, \vec {\mathcal{D}} \big(V^{-1} K^I  \big) \Big) \,.
\label{mueqn}
\end{equation}
This equation is still solved by
\begin{equation}
\mu ~=~  V^{-2}\,  \big(\coeff{1}{6} \, C^{IJK} Z_I \, Z_J \, Z_K\big) ~+~ \frac{1}{2}\, \sum_{I=1}^{N+1} V^{-1} K^{I}L_{I}  ~+~  M\,,
\label{muform2}
\end{equation}
where, once again, $M$ is a harmonic function in four dimensions. One can then use (\ref{muform2})  in (\ref{simpcovkeqn}) to simplify the right-hand side to obtain
\begin{equation}
\vec {\mathcal{D}} \times \vec \omega ~+~ V \partial_\psi \vec \omega ~=~ V \vec {\mathcal{D}} M - M\vec {\mathcal{D}} V +\frac{1}{2} \, \sum_{I=1}^{N+1} \big( K^{I} \vec {\mathcal{D}} L_{I} - L_{I}   \vec {\mathcal{D}} K^{I} \big) \,.
\label{simpomegaeqn}
\end{equation}
One can  verify that the covariant divergence (using $\vec {\cal{D}}$) generates an identity that is trivially satisfied as a consequence of  (\ref{AVreln}),  (\ref{Lorgauge}),  (\ref{muform2})  and
\begin{equation}
 \nabla^2_{(4)}  L_I ~=~  \nabla^2_{(4)}    M ~=~ 0 \,.
\label{harmonicLM}
\end{equation}
%

\subsection{Adding species on the $T^4$}

From the eleven-dimensional perspective, we are going to add extra Maxwell fields coming from three-form potentials with two legs on the $T^4$ defined by $(x^5,x^6,x^7,x^8)$ but leave the fields on the other $T^2$, defined by $(x^9,x^{10})$  unchanged and supporting only one Maxwell field, which we have labelled as $A^{(0)}$.   Thus the only non-zero components of  intersection product, $C_{IJK}$, are   
\begin{equation}
C_{0JK}  ~=~  \widehat C_{JK}  ~=~  \widehat C_{KJ} \,,
\label{RecC}
\end{equation}
for some matrix, $\widehat C_{JK} = \widehat C^{JK}$.  One can easily see that (\ref{Cconstr}) implies that, as matrices, $\widehat C^3 = \widehat C$ and so, assuming that $\widehat C$ is invertible, we have
\begin{equation}
\widehat C_{IJ} \, \widehat C_{KL} \, \delta^{JK}  ~=~  \delta_{IL}   \,.
\label{Csqone}
\end{equation}

The four-torus, $T^4$, has six independent harmonic forms and by generalizing the Ansatz (\ref{Cfield}) these forms can give rise to six vector fields in five dimensions.  However,  some  of these vector fields belong to $\Neql{2}$ gravitino multiplets and do not lie in an $\Neql{2}$ supergravity theory coupled to vector multiplets.  Indeed, the vector fields in such an $\Neql{2}$ supergravity theory  are associated with forms in $H^{(1,1)}(T^4, \IC)$   \cite{Cadavid:1995bk,Papadopoulos:1995da,Giusto:2012gt} for a suitably chosen complex structure.  We will take this complex structure to be defined by
\begin{equation}
w_1 ~=~ x^5 + i x^6 \,, \qquad   w_2 ~=~ x^7 + i x^8 \,, \qquad   w_0 ~=~ x^9 + i x^{10}  \,.
\label{cplxstr}
\end{equation}
Then the forms 
\begin{equation}
\Omega_1 ~\equiv~ \coeff{i}{2}\, d w_1 \wedge d \bar w_1 ~=~ dx^5 \wedge  dx^6 \,, \qquad \Omega_2 ~\equiv~   \coeff{i}{2}\, d w_2 \wedge d \bar w_2 ~=~ dx^7 \wedge  dx^8  \,,
\label{forms1}
\end{equation}
give rise to the other two vector fields, (apart from $A^{(0)}$),  in (\ref{Cfield}).    There are two more forms in $H^{(1,1)}(T^4, \IC)$:  
\begin{align}
\Omega_3~\equiv~ & \coeff{1}{2 \sqrt{2} }\, (d w_1 \wedge d \bar w_2 + d \bar w_1 \wedge d  w_2)  ~=~   \coeff{1}{\sqrt{2} }\, (dx^5 \wedge  dx^7 + dx^6 \wedge  dx^8)     \,,  \nonumber \\
\Omega_4 ~\equiv~ &   \coeff{i}{2\sqrt{2}}\, (d w_1 \wedge d \bar w_2 - d \bar w_1 \wedge d  w_2)   ~=~     \coeff{1}{\sqrt{2} }\, (dx^5 \wedge  dx^8 - dx^6 \wedge  dx^7)  \,.
\label{forms2}
\end{align}
This leads to an intersection matrix
\begin{equation}
\widehat C_{IJ}  ~=~  
\begin{pmatrix} 
0&1&0&0\\  1&0&0&0\\ 0&0&-1&0\\ 0&0&0&-1
\end{pmatrix}   \,,
\label{hatCform}
\end{equation}
which satisfies (\ref{Csqone}).  Indeed the normalizations in (\ref{forms2}) are set so as to satisfy  (\ref{Cconstr}) and (\ref{Csqone})  and yield canonical and uniform normalization of the Maxwell fields.

The eleven-dimensional three-form potential has therefore the form:
\begin{equation}
C^{(3)}  ~=~ A^{(0)}  \wedge dx^9 \wedge dx^{10} ~+~  \sum_{J=1}^4 \, A^{(J)} \wedge \Omega_J      \,.
\label{Cfieldnew}
\end{equation}
This is a very modest generalization of the class of theories considered in  \cite{Giusto:2012gt,Vasilakis:2012zg}.  Indeed, we can easily truncate down to  $A^{(0)}, A^{(1)}, A^{(2)},A^{(4)}$ by imposing invariance under the discrete inversion $(t,\psi,\vec y, x^5, x^7, x^9) \to -(t,\psi,\vec y, x^5, x^7, x^9)$. 

Observe that (\ref{Zreln}) now implies that the space-time metric warp factor, $Z$, is given by
\begin{equation}
Z^3   ~=~  \coeff{1}{2} \, Z_0 \, \big( \widehat C^{IJ}   Z_I \, Z_J \big)~=~   Z_0 \, \big(   Z_1 \, Z_2  - \coeff{1}{2} \, (Z_3^2  + Z_4^2) \big)\,.   
\label{Zreln2}
\end{equation}
It is also convenient to define the quadratic combination:
\begin{equation}
P   ~\equiv~ \widehat C^{IJ}   Z_I \, Z_J~=~  \big(    Z_1 \, Z_2  - \coeff{1}{2} \, (Z_3^2  + Z_4^2) \big)\,.   
\label{Pdefn}
\end{equation}

The eleven-dimensional metric in this truncation is given by
\begin{align}
 ds_{11}^2  ~=~  - Z^{-2}  (dt+k)^2  & + Z \, ds_4^2   +    
 Z  \, \Big( Z_0^{-1} \, |dw_0|^2 + P^{-1}\, Z_2 \, |dw_1|^2 + P^{-1}\, Z_1 \, |dw_2|^2 \nonumber \\
 & +  \coeff{i}{\sqrt{2}} \,P^{-1}\, Z_3 \, (dw_1 d\bar w_2 -  dw_2 d\bar w_1) -   \coeff{1}{\sqrt{2}} P^{-1}\, Z_4 \, (dw_1 d\bar w_2 + dw_2 d\bar w_1)\Big)\,.
\label{elevenmetric2}
\end{align}
%

\section{Solutions in six and ten dimensions}
\label{Sect:SixDim}

We are ultimately seeking microstate geometries  with species oscillation.  This is most easily achieved by using charge density modes on supertubes and for such geometries to be smooth we must go to six-dimensional supergravity.  Indeed, one must choose one of the vector fields to become the geometric KK fiber and the remaining $N$ vector fields become encoded in one self-dual and $N-1$ anti-self-dual tensor multiplets in the six-dimensional theory.  We will not need all the details in six dimensions because the smooth  Maxwell fields in five dimensions will become smooth tensor gauge fields in six dimensions.  The only detail we have to worry about is the singular Maxwell field coming from the supertube source in five dimensions and how this becomes smooth geometry in six dimensions. 

\subsection{Lifting to six dimensions}

The six-dimensional metric is 
\begin{align}
ds_6^2 ~=~  & 2 H^{-1} (dv+\beta) \big(du +  k  ~-~ \coeff{1}{2}\,  Z_0 \, (dv+\beta)\big) ~-~  H \, ds_4^2   \label{sixmet1} \\
 ~=~  &  (Z_0 H)^{-1} (du +  k)^2 ~-~  Z_0 \, H^{-1}  \big(dv+\beta   ~-~  Z_0^{-1} \, (du +  k)\big)^2 ~-~  H \, ds_4^2  
\,.   \label{sixmet2}
\end{align}
where the four-dimensional base is exactly that of the five-dimensional theory.  The functions $Z_0$ and $K^0$ define the geometric KK vector field and so the vector potential, $\beta$, in (\ref{sixmet1}) is 
\begin{equation}
\beta ~\equiv~ \frac{K^0}{V}  \, (d\psi ~+~ A) ~+~  \vec{\xi} \cdot  d \vec y \,,
\label{betapot}
\end{equation}
where
\begin{equation}
\vec  \nabla \times \vec \xi  ~=~ - \vec \nabla K^0 \,.
\label{xidefn1}
\end{equation}
The six-dimensional warp factor, $H$, is defined by  the quadratic  in (\ref{Pdefn})
\begin{equation}
H ~\equiv~ \sqrt{P} ~=~ \sqrt{ Z_1 \, Z_2  - \coeff{1}{2} \, (Z_3^2 + Z_4^2) } \,. \label{Hdefn}
\end{equation}

Part of the purpose of separating the $T^4$ from the $T^2$ in the M-theory description was to cleanly separate the fields that will be involved in species oscillation from the fields that would be involved in the KK uplift to six dimensions.  The fields coming from the $T^4$ in the M-theory description  will  thus be related to fields coming from the $T^4$ in the IIB compactification that leads to the six-dimensional description.

A supertube at $\vec y =0$ corresponds to having sources for $K^0$ ,  $L_I$, $I=1,\dots,N$  and $M$ of the form
\begin{equation}
K^0~\sim~ \frac{k}{r}  \,, \qquad L_I ~\sim~ \frac{Q_I}{r} \,, \qquad M ~\sim~ \frac{m_0}{r}   \,. \label{stsrc}
\end{equation}
as $r \equiv |\vec y| \to 0$ but with all other functions being regular at $r=0$.  The supertube thus runs along the $\psi$ fiber and is smeared along $v$. The  parameter $k$ is the magnetic dipole charge of the supertube,  the $Q_I$ define its electric charges and $m_0$ is its angular momentum along $\psi$.  The regularity of the six-dimensional metric then follows from the fact even though $H \sim \cO(r^{-1})$, the metric can be desingularized  by a change of radial variable and the KK fibration makes the metric transverse to the $\psi$-fiber into (an orbifold of) $\IR^4$ \cite{Lunin:2001fv,Lunin:2002iz,Bena:2008dw}.   The details of the orbifold depend upon the magnetic and geometric charges.  We will review an example below. 

Supertube regularity is not automatic:  there are still some  conditions upon the location and charges of the supertube.   Specifically, one must  ensure  a) that the coefficient of $(d\psi + A)^2$ in the metric  remains finite as $r\to 0$  and b) that there are no Dirac strings in $\omega$ originating from $r=0$. 
The latter condition  can be written in various ways; a convenient form found in \cite{Bena:2008dw} is 
\begin{equation}
\lim_{r  \to 0}\,  r \, \big[ V  \mu ~-~Z_0\,  K^0 \big] ~=~ 0 \,.
\label{stregconda}
\end{equation}
Given this condition, the finiteness of the coefficient of $(d\psi + A)^2$ in the metric is then equivalent to 
\begin{equation}
\lim_{r  \to 0}\,  r^2\, {\cal Q}  ~=~ 0  \,, \label{stregcondb}
\end{equation}
where  $\cQ$ is defined in (\ref{Qdefn}) and expressed in  simplified form in (\ref{Qreduced}). 

The two conditions, (\ref{stregconda}) and (\ref{stregcondb}), guarantee that a supertube smoothly caps off the spatial geometry, up to orbifold singularities \cite{Lunin:2001fv,Lunin:2002iz,Bena:2008dw}.

\subsection{The ten-dimensional  IIB configuration}
\label{Sect:Ten}

The six-dimensional KK uplift above is most simply related to a $T^4$ compactification of IIB supergravity.  It is relatively easy to see what the corresponding brane configuration is by compactifying and T-dualizing the M2 and M5 configurations described in Section \ref{Sect:AddVMs}.  

If one compactifies on $x^{10}$  then the M2-branes of species ``0'' become F1-strings wrapping $x^9$ and all the other M2-branes become D2-branes. The M5-dipole charge of species ``0'' becomes an NS5-dipole charge wrapping  $\psi, x^5,\dots, x^{8}$, where $\psi$ is the GH fiber, while all the other M5-dipole charges become D4-dipole charges.  Performing the T-duality on $x^9$ converts the F1-strings and NS5-branes to momentum and KK monopole charges  respectively, and all the D2-charges become D3-charges intersecting on a common $x^9$ and wrapping the $T^4$ defined by  $(x^5,\dots, x^{8})$ in exactly the same manner as the original M2-charges.  Similarly, the D4-dipole charges become D3-dipole charges intersecting on the common $\psi$-circle and wrapping the $T^4$ defined by  $(x^5,\dots, x^{8})$ in exactly the same manner as the original M5 dipole charges.

Thus we obtain a IIB configuration in which all the M2 species have become either D3-branes or momentum. One can perform two more T-dualities on the $T^4$, for example along $(x^5,x^6)$, thereby converting two of the sets of D3-charges into D1- and D5-charges. This T-duality changes the other D3-charge species into D3-charges with different orientation, and introduces additional NS-NS B-fields on the $T^4$. Thus one can obtain the canonical D1-D5-P system decorated with additional D3-charges.  This last T-duality does not affect the space-time metric and its regularity.

There is another way to dualize our configurations to magnetubes with D1 and D5 charges. If one starts from a magnetube whose D1 and D5 charges oscillate into F1 and NS5 charges \cite{Mathur:2013nja}, there is a duality sequence, given by equation (3.1) in \cite{Giusto:2012gt}, that takes this solution into a solution of the type we construct here but where the species corresponding to $A^3$ and $Z_3$ is absent. Hence, if one applies the inverse of this duality sequence on a solution with  $A^3=0= Z_3$ one obtains the magnetube of  \cite{Mathur:2013nja}. If one applies the inverse duality sequence to a more general solution, where the third species is also present, the resulting magnetube will have two extra charges, corresponding to D3 branes with two legs along the $T^4$ and one along the common D1-D5 circle.

\section{The non-oscillating template}
\label{Sect:Temp}

We now describe in detail a simple example of a smooth magnetube geometry, which we will construct by putting a single supertube (with no species oscillation) on the $V=0$ surface in the simplest ambipolar Gibbons-Hawking space. The discussion here follows closely previous work on supertubes in such backgrounds, but there is some novelty in dealing with putting the supertube on the $V=0$ surface. Our aim in this section is to show that the conditions for smoothness of the supertube are unchanged by placing it on the $V=0$ surface, and hence translate directly in the conditions for smoothness of a magnetube. The non-oscillating solution provides a simple context to deal with these issues and set up the framework before we move on to introduce species oscillation. We will see later that for the solutions with species oscillation we can choose the charge densities such that the angular momentum density remains constant, and then the structure of the metric in the oscillating solutio
n is almost identical to that of the non-oscillating solution.

\subsection{$AdS_3 \times S^2$}

We start from the simplest solution with a non-trivial critical surface, a two centered metric. We first review the background solution before we introduce the supertube; we consider a particularly simple example in which  the  five-dimensional geometry is simply $AdS_3 \times S^2$ \cite{Denef:2007yt,deBoer:2008fk,Bena:2010gg}. This solution is axisymmetric and so it is more convenient to write the four-dimensional base in terms of cylindrical polars on the flat $\IR^3$ slices: 
\begin{equation}
ds_4^2 ~=~ V^{-1} \, \big( d\psi + A)^2  ~+~ V\, (d\rho^2 + dz^2 + \rho^2 d \phi^2) \,. \label{GHpolar}
\end{equation}
For 
simplicity we take the two GH charges to be equal and opposite, so
\begin{equation}
 V ~=~  \frac{q}{r_+} ~-~  \frac{q}{r_-} \,,
\label{Vform}
\end{equation}
where
\begin{equation}
 r_\pm ~\equiv~  \sqrt{\rho^2 ~+~ (z\mp a)^2} 
\label{rpmdefn}
\end{equation}
for some parameter $a$. In this simple two-centered solution, the solution will have the classic form  of Section \ref{Sect:classic} with two vector multiplets, except we will use the labeling   $\{I, J, K\} = \{0, 1, 2\}$  instead of the more common usage ($\{I, J, K\} = \{1, 2, 3\}$).  

We initially take the magnetic flux functions, $K^I$, to have the simple form
\begin{equation}
K^0 ~=~  K^1 ~=~ K^2 ~=~ K =  \frac{k}{r_+} ~+~  \frac{k}{r_-} \,, 
\label{Ktwo}
\end{equation}
where $r \equiv \sqrt{\rho^2 + z^2}$.  The electric source functions, $L_I$, are set to
\begin{equation}
 L_0 ~=~  L_1 ~=~ L_2 ~=~  L= -\frac{k^2}{q} \Big( \frac{1}{r_+} ~-~  \frac{1}{r_-} \Big)  
\label{Ltwo}
\end{equation}
The coefficients of the $r_\pm^{-1}$ terms in the $L_I$ have been fixed by the usual regularity condition for bubbled geometries in five-dimensions: $Z_I$ should be finite at $r_\pm =0$. 

The harmonic function $M$ in the angular momentum is
\begin{equation}
M ~=~  m_\infty~+~  \frac{k^3}{2\, q^2} \Big( \frac{1}{r_+} ~+~  \frac{1}{r_-} \Big) \,.
\end{equation}
Again the coefficients of the $r_\pm^{-1}$ terms in  $M$ have been fixed by the usual five-dimensional regularity conditions for bubbled geometries: $\mu$ must be finite at $r_\pm =0$.  The  parameter $m_\infty$ is introduced to ensure smoothness and lack of Dirac strings. Finally, the $3$-dimensional angular-momentum vector, $\omega$,  is determined using (\ref{omegeqn}):
\begin{equation}
\omega ~=~   - \frac{2\, k^3}{q \,a} \,\Big( \frac{\rho^2 + (z-a +r_+)  (z+a - r_-)}{r_+ r_-}  \Big) \, d \phi\,. 
\end{equation}
To make sure that there are no CTC's near $r_\pm =0$ and to remove Dirac strings, we must impose the five-dimensional bubble equations: $\mu(r_\pm =0) =0$.  For this two-centered solution this requires only that 
\begin{equation}
m_\infty = - \frac{2 \, k^3}{a\, q^2} .
\end{equation}

To demonstrate that the resulting  five-dimensional ambi-polar geometry is just $AdS_3 \times S^2$, we introduce bipolar coordinates centered on $r_\pm = 0$:
\begin{equation}
 z =  a\, \cosh 2\xi \,\cos \theta \,, \qquad  \rho =  a\, \sinh 2\xi \, \sin \theta \,, \qquad
 \xi \ge 0\,, \ \ 0 \le \theta \le \pi \,.
  \label{coordsa}
 \end{equation}
In particular, one has 
\begin{equation}
r_\pm   ~=~  a\, (\cosh 2\xi  \mp \cos \theta) \,.
  \label{rpmform}
 \end{equation}
Rescaling and shifting the remaining variables according to
\begin{equation}
 \tau ~\equiv~   \coeff{a\, q}{8\, k^3}\,  u  \,, \qquad \varphi_1 ~\equiv~   \coeff{1}{2\,q} \, \psi -
 \coeff{a\, q}{8\, k^3}\,  u   \,, \qquad \varphi_2 ~\equiv~ \phi -   \coeff{1}{2\,q} \, \psi +
 \coeff{a\, q}{4\, k^3}\,  u   \,,
 \label{coordsb}
 \end{equation}
the five-dimensional metric  then takes the standard {\it global} $AdS_3 \times S^2$ form:
\begin{equation}
ds_5^2 ~\equiv~ R_1^2 \big[ - \cosh^2\xi \,  d\tau^2 + d\xi^2 +  \sinh^2 \xi \, d\varphi_1^2 \big] ~+~  R_2^2 \big[   d \theta ^2 + \sin^2\theta  \, d\varphi_2^2 \big]  \,,
 \label{AdS3S2}
 \end{equation}
with
\begin{equation}
  R_1~=~  2 R_2 ~=~ 4 k \,.
 \label{Radii}
 \end{equation}
One can further check that the $v$-fiber in (\ref{sixmet2}) adds a Hopf fiber to the $S^2$ making the metric that of {\it global} $AdS_3 \times (S^3/\ZZ_{2k})$:
\begin{equation}
ds_6^2 ~=~ R_1^2 \big[  \cosh^2\xi \,  d\tau^2 - d\xi^2 -  \sinh^2 \xi \, d\varphi_1^2 \big] ~-~  R_2^2 \big[   d \theta ^2 + \sin^2\theta  \, d\varphi_2^2 + \big(\coeff{1}{2k} \, dv -  \cos \theta \, d\varphi_2 \big)^2\big]  \,.
 \label{AdS3S3}
 \end{equation}

One  should note that an observer whose world-line has tangent, $\frac{\partial}{\partial u}$, {\it i.e.} the observer is fixed in the GH spatial base, follows a curve with 
\begin{equation}
\frac{d \varphi_1}{ d\tau}  ~=~ 1 \,, \qquad \frac{d \varphi_2}{ d \tau}  ~=~  2   \,,
 \label{velocties}
 \end{equation}
which means that the proper $4$-velocity has norm: $-4k^2 \cos^2 \theta$.  Thus this GH-stationary observer follows a time-like curve everywhere except on the equator ($\theta = \frac{\pi}{2}$) of the $S^2$  where  the observer's trajectory becomes null. This illustrates the general discussion in Section \ref{Sect:Merge}:  in patches that are smooth across the critical surface $V=0$,  the GH-stationary observers are being boosted from time-like to null trajectories as $V\to 0$.

\subsection{Adding the supertube/magnetube}

We will now add a supertube sitting on the $V=0$ surface at $\rho = z = 0$ to the five-dimensional geometry \eqref{AdS3S2}. From the  discussion above we see that this corresponds to  putting a supertube in $AdS_3 \times (S^3/\ZZ_{2k})$ so that it spirals around the equator of the $S^2$ while sitting a fixed point ($\xi =0$) in $AdS_3$. 

Technically, since the supertube is located on top of a $V=0$ surface, it is a magnetube. However, we will continue to describe it as a supertube because the mathematical construction and regularity conditions remain unchanged.

We introduce the supertube by taking the magnetic flux functions $K^I$ to have the simple form
\begin{equation}
K^0 ~=~  \frac{k}{r_+} ~+~  \frac{k}{r_-} ~+~  \frac{\alpha}{r}  \,, \qquad  K^1 ~=~ K^2 ~=~  \frac{k}{r_+} ~+~  \frac{k}{r_-} \,, 
\label{Kform}
\end{equation}
where $r \equiv \sqrt{\rho^2 + z^2}$.  The electric source functions $L_I$ are set to
\begin{equation}
 L_0 ~=~  -\frac{k^2}{q} \Big( \frac{1}{r_+} ~-~  \frac{1}{r_-} \Big)  \,, \qquad
  L_I  ~=~  -\frac{k^2}{q} \Big( \frac{1}{r_+} ~-~  \frac{1}{r_-} \Big) ~+~  \frac{\beta_I}{r} \,, \quad I=1,2\,. 
  \label{Lform}
\end{equation}
The coefficients of the $r_\pm^{-1}$ terms in the $L_I$ are as before.  The sources at $r =0$ in $K^0$, $L_1$ and $L_2$ correspond to those of a supertube with dipole charge $\alpha$ and electric charges $\beta_1$ and $\beta_2$.  We want to demonstrate that for appropriate choices of these parameters the addition of this supertube leads to a smooth six-dimensional geometry that  asymptotically approaches an  $AdS_3 \times  (S^3/\ZZ_{2k+\alpha})$ geometry.

The harmonic function $M$ in the angular momentum is now
\begin{equation}
M ~=~  m_\infty~+~  \frac{k^3}{2\, q^2} \Big( \frac{1}{r_+} ~+~  \frac{1}{r_-} \Big) ~+~  \frac{m_0}{r} \,.
  \label{Mform}
\end{equation}
Again the coefficients of the $r_\pm^{-1}$ terms in  $M$ are as before.  The parameter $m_0$ represents the angular momentum of the supertube while the parameter $m_\infty$ is introduced to ensure smoothness and lack of Dirac strings.

Finally, the $3$-dimensional angular-momentum vector, $\omega$,  is determined using (\ref{omegeqn}) and we now find
\begin{align}
\omega ~=~  & - \bigg[ \, \frac{2\, k^3}{q \,a} \,\Big( \frac{\rho^2 + (z-a +r_+)  (z+a - r_-)}{r_+ r_-}  \Big) \nonumber \\ 
 & ~+~  \frac{1}{a}\,\Big( m_0 \, q  + \frac{\alpha\, k^2}{2\, q}  \Big) \, \Big( \frac{\rho^2 + (z-a +r_+)  (z -r)}{r\,  r_+}  
~+~ \frac{\rho^2 +  (z+ r)(z+a -r_-) }{r\,  r_-} \Big)   \bigg]\, d \phi\,. 
\label{omres}
\end{align}
%

\subsection{Regularity near sources}

To make sure that there are no CTCs near $r_\pm =0$ and to remove Dirac strings, we must impose the five-dimensional bubble equations: $\mu(r_\pm =0) =0$.  This now imposes two constraints:
\begin{equation}
\beta_1 ~=~ -\beta_2 ~=~ -\beta \,,
\label{constr1}
\end{equation}
and
\begin{equation}
m_\infty ~+~ \frac{m_0}{a}  ~+~ \frac{2 \, k^3}{a\, q^2}  ~+~ \frac{\alpha \, k^2}{2\, a\, q^2}     ~=~ 0  \,.
\label{constr2}
\end{equation}
The first condition is crucial for the existence of the magnetube. The fact that the M2 charges are equal and opposite ensures their cancellation in the Killing spinor equations, and therefore the fact that the Killing spinors of the magnetube are those of M5 branes and momentum. The remaining parameters of the solution are $\alpha, \beta$ and $m_0$.

With the magnetube in place, we must now also consider the regularity conditions at $r=0$. The condition that there are no Dirac strings at $r=0$ is given by (\ref{stregconda}).  Alternatively,  it can also be read off from (\ref{omegeqn}) by simply making sure that $\vec \omega$ has no sources of the form a constant multiple of $\vec \nabla ( \frac{1}{r})$.  Using the fact that $V$ and $L_3$ vanish at $r=0$, one can easily see that this condition requires that 
\begin{equation}
\lim_{r  \to 0}\,  r \, (K^1 \, L_1 + K^2 \, L_2)~=~ 0 \,.
\label{stregcondc}
\end{equation}
Since $K^1 = K^2$, the absence of Dirac strings at the magnetube is already guaranteed by (\ref{constr1}).  This is, of course, to be expected since a Dirac string must have two ends and we have eliminated them everywhere else.

The other supertube regularity condition (\ref{stregcondb}) is easily computed from (\ref{Qreduced}) particularly since $V=0$ and $L_3 =0$ at $r=0$.  One finds that this condition is satisfied if 
\begin{equation}
m_0 ~=~ -\frac{\beta^2}{2\alpha} \,,
\label{constr3}
\end{equation}
which is the standard relationship between the charges and angular momentum of a supertube. The free parameters are then just the charges $\alpha, \beta$, as is standard for a supertube.

\subsection{Geometry of the regular magnetube}

For the later discussion of the oscillating solution, it is useful to summarize here the functions appearing in the six-dimensional metric (\ref{sixmet1}) for the non-oscilating magnetube solution and to note explicitly where the parameter $\beta$ enters into the solution. The KK electric function $Z_0$ is independent of the magnetube parameters, 
\begin{equation}
Z_0 = \frac{K^1 K^2}{V} + L_0 = \frac{K^2}{V} + L,
\end{equation}
where $K$, $L$ are given by (\ref{Ktwo}) and (\ref{Ltwo}). The other electric functions are 
\begin{equation} \label{Z12}
Z_{1,2} = Z_0 + \frac{\alpha}{r} \frac{K}{V}  \pm \frac{\beta}{r},
\end{equation}
so the warp factor $H$ is 
\begin{equation}
\label{Htemp}
H^2 = \left( Z_0 + \frac{\alpha}{r} \frac{K}{V} \right)^2 - \frac{\beta^2}{r^2} \,.
\end{equation}
As noted above, (\ref{constr1}) implies a cancellation that makes $\mu$ depend upon  $\beta$ only through $M$ and the condition  (\ref{constr3}):
\begin{equation}
\mu = \frac{K^3}{V^2} + \frac{\alpha}{r} \frac{K^2}{V^2} - \frac{3k^2}{2q^2} K - \frac{k^2}{2 q^2} \frac{\alpha}{r} + M,  
\end{equation}
where $M$ is given by  (\ref{Mform}) with the parameters fixed by (\ref{constr2}) and (\ref{constr3}). Thus, the electric charge parameter, $\beta$, only enters the metric (\ref{sixmet1}) directly through the warp factor, $H$, and appears indirectly in the one-form $k$ through the dependence of $m_0$ on $\beta$ from (\ref{constr3}). Of course, $\beta$ will also appear directly in the gauge fields $A^{(1)}$ and $A^{(2)}$. 

\subsection{$AdS_3 \times S^3$ asymptotics at infinity}

One can use the spherical bipolar coordinates (\ref{coordsa}) to study the asymptotic form of our solution. The warp factor $H$ will behave as 
\begin{equation}
H \approx \frac{k (2k+\alpha)}{q a \cos \theta} + \mathcal{O}(e^{-4 \xi}). 
\label{Hasymp}
\end{equation}
To obtain the leading asymptotics, we only need to consider the first term.  As $\xi \to \infty$, the leading-order behaviour of the metric is 
\begin{align}
ds_6^2 ~\sim~ & 8k(2k+\alpha) \, \Big[\,\frac{a^2 q^2  }{32 k^2 \, (2k + \alpha)\Delta}\,e^{2\xi} \, du^2  ~-~ d\xi^2 ~+~  \frac{\Delta}{8k^2 \, (2k + \alpha)}  \, e^{2\xi} \,  d  \hat\varphi_1^2 \, \Big]  \nonumber \\ &
~-~ 2k(2k+\alpha) \,  \Big[   d \theta ^2 + \sin^2\theta  \, d\hat \varphi_2^2 + \big(\coeff{1}{2k+\alpha} \, dv -  \cos \theta \, d\hat \varphi_2 \big)^2\Big]  \,,
 \label{asympmet1}
 \end{align}
where 
\begin{equation}
\hat \varphi_1  ~\equiv~   \frac{1}{2\, q} \, \psi  ~-~ \frac{a \, q \,  }{2\, \Delta}   \,   u   \,, \qquad \hat \varphi_2  ~\equiv~    \phi  ~-~  \frac{\Delta }{4\, q \, k^2 (2k+\alpha)} \, \psi~+~ \frac{a \, q  }{2\, k^2\,(2k+\alpha)}   \,   u    \,
 \label{coordsc}
 \end{equation}
and
\begin{equation}
\Delta ~\equiv~    -2aq^2 m_\infty = k^2 (4 k+ \alpha) - \frac{ q^2 \beta^2}{\alpha} \,.
 \label{Deltadefn}
 \end{equation}
Observe that  (\ref{coordsc})  reduces to  (\ref{coordsb}) and that  (\ref{asympmet1}) matches the asymptotic form of  (\ref{AdS3S3}) if one first sets $\beta =0$ and then sends $\alpha \to 0$.  

One can  absorb the constants in first part of (\ref{asympmet1}) by rescaling $u$ and making a constant shift in $\xi$.  The asymptotic form of the metric is thus that of $AdS_3 \times (S^3/\ZZ_{2k+\alpha})$ and is therefore smooth, provided $\Delta >0$, or
\begin{equation}
\beta^2 ~<~      \frac{\alpha\,k^2\,(4 k+\alpha)} {q^2}  \,.
 \label{bound1}
 \end{equation}
One can, in fact, allow equality here but then the metric becomes asymptotic to some form of null wave in which only the $d\psi\, du$ term survives at infinity.  If $\Delta <0$ then the $\psi$-circles become closed time-like curves.  We assume that (\ref{bound1}) is true.

The leading-order asymptotics of the matter fields can similarly be calculated; the two non-KK scalar fields are
\begin{equation}
X^1 ~=~  \bigg(\frac{Z_0 Z_2}{Z_1^2}\bigg)^{1/3}, \qquad X^2  ~=~   \bigg(\frac{Z_0 Z_1}{Z_2^2}\bigg)^{1/3} ,
\end{equation}
so asymptotically 
\begin{equation}
X^1 = X^2 =  \left( 1 + \frac{\alpha}{2k} \right)^{-1/3} + \mathcal{O}(e^{-4\xi}).
\end{equation}
The background value of the scalar fields is rescaled along with the $AdS_3$ radius of curvature, and there is no sub-leading  part at order $e^{-2\xi}$, so there is no vev for the dual operators. The two non-KK vector fields are 
\begin{equation}
A^{(I)} = - Z_{I}^{-1} (dt+k) + \frac{K^I}{V} (d\psi + A) + \vec{\xi}^{(I)} \cdot d\vec{y} \,, \quad I=1,2 \,,
\end{equation}
where $\vec{\nabla} \times \vec{\xi}^{(I)} = - \vec{\nabla} K^{(I)}$, and $Z_I$ are given by (\ref{Z12}). In the pure $AdS_3$ background, $A^{(I)} = -2k \cos \theta d\varphi_2$. The leading asymptotics in the presence of the magnetube are  
\begin{equation}
A^{(I)}  = -2k \cos \theta d \hat \varphi_2 \mp \frac{\beta}{(2k+\alpha)} d\psi + \mathcal{O}(e^{-2\xi})\,, \quad I=1,2 \,,
\label{deltaA} 
\end{equation}
so we get the same component on the sphere, consistent with the fact that the sphere's radius of curvature is unchanged in the metric \eqref{asympmet1}. The second term is locally pure gauge, but as $\psi$ is a compact direction it introduces a Wilson line.  

In $AdS_3$, the asymptotic behavior of a massless gauge field is $A_\mu \sim A_\mu^{(0)} + j_\mu \ln r$, where $A_\mu^{(0)}$ is interpreted as the boundary gauge field and $j_\mu$ is the vev of the boundary conserved current. The absence of a logarithmic term in $r$ (a linear term in $\xi$) in the expansion (\ref{deltaA}) thus indicates that the introduction of the magnetube does not produce any charge density from the point of view of the dual CFT; the only effect of $\beta$ is to turn on a Wilson line along the $\psi$ circle.

\subsection{Global regularity}

As we remarked earlier, the two conditions (\ref{stregconda}) and (\ref{stregcondb}) guarantee that a supertube smoothly caps off the spatial geometry \cite{Lunin:2001fv,Lunin:2002iz,Bena:2008dw} and so, in principle, (\ref{constr1}), (\ref{constr2}) and (\ref{constr3}) guarantee that the supertube is smooth in six dimensions.  However, we actually have a magnetube: that is, a supertube  located at the critical ($V=0$) surface. While this should not affect the arguments of \cite{Lunin:2001fv,Lunin:2002iz}, we will now examine the metric in more detail to confirm that the magnetube limit of the supertube does not add any further subtleties.

Even though the metric appears to be singular at the critical surfaces ($V=0$), it is well known that it is actually smooth across such surfaces (see, for example, \cite{Bena:2007kg}).  There is also an apparent singularity at $r=0$ but this is resolved by the standard change of coordinate that shows that the supertube is smooth in six dimensions \cite{Lunin:2001fv,Lunin:2002iz,Bena:2008dw}. In the solutions we are considering, these two apparent singularities coincide and while the methods of resolution of the two types of apparent singularity are not expected to interfere with each other,  it is important to make sure.

Expanding   (\ref{sixmet2}) in spherical polar coordinates around $r=0$, one obtains the leading order behavior:
\begin{equation}
ds_6^2 ~\sim~ - \frac{2\,k\,\alpha}{a} \, \Big[\,  \frac{dr^2}{r} ~+~ r\, \big(d\theta^2 ~+~ \sin^2 \theta\, \tilde d\varphi_2^2  ~+~(\coeff{1}{\alpha} dv-  \cos \theta\, d \tilde \varphi_2)^2 \big)     
~-~ \coeff{1}{\cQ_0}  \, du^2 ~+~\coeff{a^4 \,q^2\, \cQ_0 }{4\, k^4 \,\alpha^2}   \, d \tilde \varphi_1^2 \Big]\,,
 \label{asympmet2}
 \end{equation}
where
\begin{equation}
\tilde \varphi_1  ~\equiv~   (\coeff{1}{2\,q }\, \psi - \phi)  - \coeff{2\, k^2\,\alpha  }{q \,a^2}  \, u   \,, \qquad \tilde \varphi_2  ~\equiv~    \coeff{1}{2} (\coeff{1}{2q}  \psi +  \phi) - \coeff{q^2 \, \beta^2 }{2\,k^2\,\alpha^2}  (\coeff{1}{2q}   \psi - \phi)    \,.
 \label{coordsd}
 \end{equation}
and 
\begin{equation}
\cQ_0  ~\equiv~   \lim_{r\to 0} \, (r \, \cQ) ~=~   \frac{4\,k^2 }{q^2 \, a^3 \,\alpha} \,\big( (2k+\alpha)\,k^2\,\alpha^2  ~+~ (2k-\alpha)\, q^2 \,\beta^2 \big)\,.
 \label{cQ0defn}
 \end{equation}
Provided that $\cQ_0>0$, the metric (\ref{asympmet2}) is manifestly a time and $S^1$ fibration defined by $(u,  \tilde \varphi_1)$ over a four-dimensional spatial base defined by $(r,\theta,\tilde \varphi_2,v)$.  The apparent singularity at $r=0$ in this spatial base is resolved in the usual manner by changing to a new coordinate, $R$, defined by $r = \frac{1}{4} R^2$.  Thus this spatial base is an orbifold of $\IR^4$, and the whole metric is smooth up to such orbifold singularities.   Indeed if one restricts to slices of constant $(u,  \tilde \varphi_1)$, then one has $d \phi  = \frac{1}{2\,q }\, d\psi$ and the four dimensional base metric becomes
\begin{equation}
ds_4^2 ~\sim~ - \frac{2\,k\,\alpha}{a} \, \Big[\,  dR^2~+~ \coeff{1}{4} \,R^2\, \big(d\theta^2 ~+~ \sin^2 \theta\,  (\coeff{1}{2\,q }\, d\psi)^2  ~+~(\coeff{1}{\alpha} dv-  \cos \theta\, (\coeff{1}{2\,q }\, d\psi) \,)^2 \big)  \,\Big]   \,.
 \label{basemet1}
 \end{equation}
If one remembers that $\psi$ has period $4 \pi$ then one sees that this is precisely the metric of flat $\IR^4$ for $q = \alpha =1$.  Other integer values of $\alpha$ and $q$ thus lead to orbifold singularities.

Smoothness around $r=0$ this requires $\cQ_0>0$.  If $2k >\alpha$ then this is manifest, but if not one can use  (\ref{bound1}) to replace $\beta^2$ and conclude that  
\begin{equation}
\cQ_0  ~>~   \frac{4\,k^4 }{q^2 \, a^3} \,\big( (2k+\alpha)\,\alpha ~+~(2k-\alpha)\, (4 k+\alpha) \big)~=~  \frac{32\,k^6 }{q^2 \, a^3}\,,
 \label{bound2}
 \end{equation}
which establishes positivity.

Global regularity of the solution also requires that the functions $HV$, $Z_1 Z_2^{-1}$ and $Z_0 H^{-1}$ are globally positive and well-behaved, except for possible singularities allowed by supertubes.  In particular, this means that all the functions $Z_I V$ must be globally positive.  This is trivial to see for $Z_0 V$ since it may be explicitly written as 
\begin{equation}
Z_0 \, V   ~=~   \frac{4\,k^2}{r_+ \, r_- } \,.
 \label{Z0V}
 \end{equation}
Similarly one can write 
\begin{align}
Z_I \, V   ~=~ &  \frac{1}{r \, r_+ \, r_- } \,\big[  4 k^2 \, r ~+~   k\, \alpha \, (r_+  + r_-)~\pm~  q \, \beta \, (r_+ - r_-)  \big] \nonumber \\
 ~\ge~ &  \frac{1}{r \, r_+ \, r_- } \,\big[  4 k^2 \, r ~+~   k\, \alpha \, (r_+  + r_-)~-~  |q \, \beta \, (r_+ - r_-)|  \big]  \nonumber \\
  ~>~ &  \frac{1}{r \, r_+ \, r_- } \,\big[  4 k^2 \, r ~+~   k\, \alpha \, (r_+  + r_-)~-~ k\,(2k + \alpha) \,|\, (r_+ - r_-)|  \big]  \,.
 \label{ZIV}
 \end{align}
where we have used the fact that (\ref{bound1}) implies $ |q \, \beta | < k(2k + \alpha)$.  Suppose $(r_+ - r_-)>0$, then we may write this as    
\begin{equation}
Z_I \, V    ~>~  \frac{1}{r \, r_+ \, r_- } \,\big[   2 \, k^2\,  (r  + r_-  -a ) ~+~   2 \, k^2\,  (r  + a -  r_+  ) +2\, k\,\alpha\, r_-  \big] ~\ge~ 0 \,,
 \label{bound3}
 \end{equation}
where the positivity follows from the triangle inequality.  One can easily permute this to obtain the result for $(r_+ - r_-)<0$.  

It is worth noting that this bound is saturated if one takes $r_-=0$ and thus $r=a$ and $r_+ =2a$.  This means that the weaker bound   (compared to (\ref{bound1})) 
\begin{equation}
|q \, \beta | ~<~  k(2k + \alpha)   \,,
 \label{bound4}
 \end{equation}
could also have been deduced from the positivity of $Z_I \, V$ at either $r_-=0$ or $r_+=0$.

Thus the basic bound (\ref{bound1}) on the supertube charge guarantees that the metric is non-singular. 

Finally, one must make a global check for CTC's and for this we have to resort to numerical methods.  One needs to verify that the metric is stably causal, that is, the metric has a global time function. This condition is simply \cite{Berglund:2005vb}:
\begin{equation}
- g^{\mu\nu} \partial_{\mu}t \, \partial_{\nu} t = - g^{tt} =  ((Z_0 V) (Z_1 V)  (Z_2 V) )^{-1/3}  (\cQ -  \omega^2) > 0\,,
\label{stabcausal}
\end{equation}
where $\omega$ is squared using the $\IR^3$ metric.  Thus we need to verify the global positivity of $(\cQ -  \omega^2)$.

\begin{figure}[t]
\centering
\includegraphics[width=12cm]{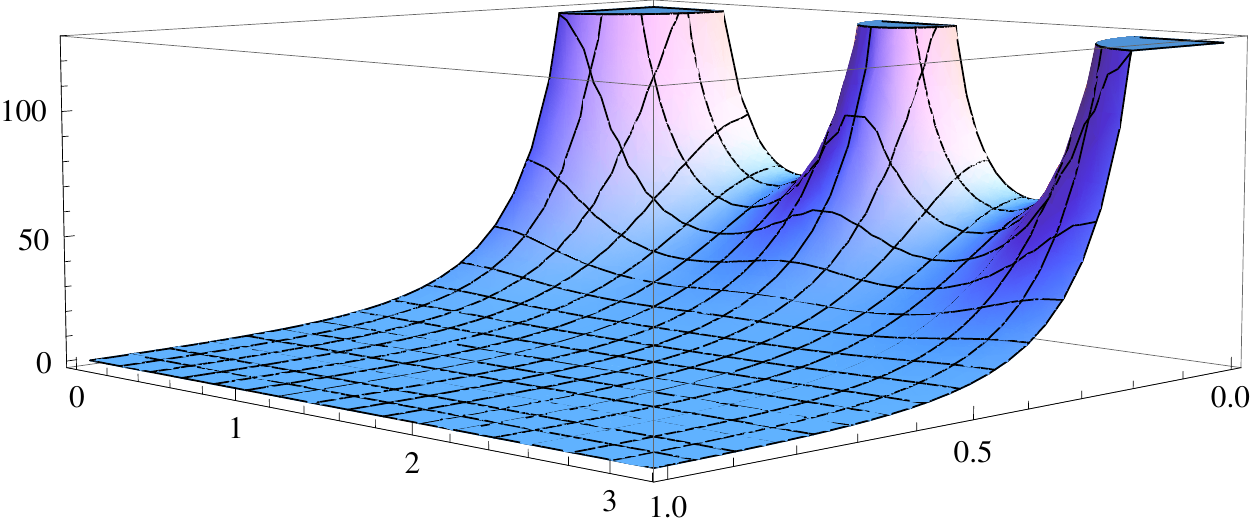}
\caption{Plot of $(\cQ -  \omega^2)$ for $0 < \xi < 1$, $0< \theta < \pi$ with parameters $q=k=\alpha = a =1$ and $\beta =2$.  These choices obey (\ref{bound1}) and the plot shows that $(\cQ -  \omega^2) >0$.}
\label{Qpic}
\end{figure}

For large values of $r$, one finds 
\begin{equation}
  \lim_{r\to \infty} \,r^3 \,  (\cQ-\omega^2) ~=~  \lim_{r\to \infty} \,r^3 \,  (\cQ)   ~=~ \frac{ 4\, k^2\,(2k + \alpha)\, \Delta}{a \, \alpha \, q^2} \,,
 \label{cQinfty}
 \end{equation}
where $\Delta$ is defined by (\ref{Deltadefn}) and hence  (\ref{bound1}) guarantees that $(\cQ -  \omega^2)$ is positive as one goes to infinity. 
Beyond this, we have examined this quantity in several  examples of solutions satisfying (\ref{bound1}) and found them to be stably causal.  A typical example is shown in Fig. \ref{Qpic}.

\section{Species oscillation}
\label{Sect:SOsoln}

Our primary contention in this paper is that species oscillation can be done on any $V=0$ surface in any microstate geometry based upon a generic GH base. However, the essential Green's functions for such generic microstate geometries are not explicitly known.  On the other hand, the introduction of species oscillation involves a localized source and the  two-centered  solution considered in Section \ref{Sect:Temp} provides an excellent ``local model'' of a typical $V=0$ surface in a generic microstate geometry.  Moreover, the Green's functions for this relatively simple system are known \cite{Bena:2010gg} and so the solutions can be constructed explicitly.  

Given that species oscillation can be implemented, in a very straightforward manner, to produce new classes of microstate geometries  within such a ``local model,'' we believe that the generalization to any microstate geometry should present no difficultly apart from the fact that explicit analytic examples may not be available. 

\subsection{Some simple oscillating solutions}

We will thus construct the simplest example of a non-trivial solution with species oscillation: We add a magnetube with an oscillating charge distribution to the simple two-centre background from the previous section.  The first step will be to allow the electric charge sources in (\ref{Lform}) to be $\psi$-dependent much as in \cite{Bena:2010gg}, however here these charges will oscillate into other species of M2-brane charge.    To enable this species oscillation we  add two vector multiplets as discussed in section \ref{Sect:AddVMs} but only give allow them to have electric charges.  

We therefore take the GH base with $V$ still given by (\ref{Vform}) and keep the same magnetic sources as in  (\ref{Kform}):   %
\begin{equation}
K^0 ~=~  \frac{k}{r_+} ~+~  \frac{k}{r_-} ~+~  \frac{\alpha}{r}  \,, \qquad  K^1 ~=~ K^2 ~=~  \frac{k}{r_+} ~+~  \frac{k}{r_-} \,, \qquad  K^3 ~=~ K^4 ~=~ 0\,. 
\label{sameKform}
\end{equation}
The electric source functions, $L_I$, are set to:
\begin{align}
 L_0 & ~=~  -\frac{k^2}{q} \Big( \frac{1}{r_+} ~-~  \frac{1}{r_-} \Big)  \,,\nonumber \\ 
  L_I  & ~=~  -\frac{k^2}{q} \Big( \frac{1}{r_+} ~-~  \frac{1}{r_-} \Big) ~+~  \lambda_I \,, \quad I=1,2 \,;\nonumber \\ 
   L_I & ~=~ \lambda_I \,, \quad I=3,4   \,, 
  \label{newLform}
\end{align}
where the oscillating charge fields, $\lambda_I$ = $\lambda_I(\psi,\vec y)$, are harmonic functions in four dimensions 
\begin{equation}
\nabla^2_{(4)} \lambda_I  ~=~  0\,,
\label{harmlam}
\end{equation}
that will be required to have no more than an $\cO(r^{-1})$ singularity as $r \to 0$: 
\begin{equation}
 \lambda_I(\psi,\vec y)  ~\sim~  \frac{\rho_I(\psi)}{r}   \,,
\label{lamsrc}
\end{equation}
for some charge-density functions, $\rho_I(\psi)$. As before, this corresponds to a supertube that wraps around the $\psi$ circle and now carries four kinds of electric charges, $\rho_I(\psi)$, as well as the magnetic dipole charge $\alpha$. We will assume that the integral over $\psi$ of the charge-density functions $\rho_I(\psi)$ vanishes, so that (as we will show later for the asymptotic charges) the supertube is carrying no net electric charge; the electric charge oscillates  between the different species as we go around the $\psi$ circle.  This is the key difference from previous studies with varying charge density  \cite{Bena:2010gg}, and is possible because we take the supertube on the $V=0$ surface. 

Since $K^3 = K^4 \equiv 0$, it follows that 
\begin{equation}
Z_I(\psi,\vec y)  ~=~  \lambda_I \,, \quad I=3,4 \,.
\label{Z34simp}
\end{equation}
The function, $M$, must now incorporate the possibility of a $\psi$-dependent source at $r=0$:
\begin{equation}
M ~=~  m_\infty~+~  \frac{k^3}{2\, q^2} \Big( \frac{1}{r_+} ~+~  \frac{1}{r_-} \Big) ~+~   \eta(\psi,\vec y) \,.
  \label{newMform}
\end{equation}
where  $\eta(\psi,\vec y)$ is harmonic and with no more than an $\cO(r^{-1})$ singularity as $r \to 0$: 
\begin{equation}
 \eta(\psi,\vec y)  ~\sim~  \frac{\hat \rho (\psi)}{r}   \,.
\label{etasrc}
\end{equation}
%

\subsection{Regularity}

It was shown in \cite{Bena:2010gg} that for the $\psi$-dependent solutions we are considering here, supertube regularity can be analyzed locally in $\psi$ and that regularity is, once again, ensured by imposing (\ref{stregconda}) and (\ref{stregcondb}).  As before, (\ref{stregconda}) guarantees the absence  of Dirac strings and, equivalently,  this relevant condition can be obtained from (\ref{simpomegaeqn}) by requiring that $\vec \omega$ has no sources of the form of a constant multiple of $\vec \nabla \frac{1}{r}$.  Using the fact that $K^1 \equiv  K^2$, $K^3 = K^4 \equiv 0$ and  $V$ and $L_0$ vanish at $r=0$, this Dirac string condition implies that the singularities in $L_1$ and $L_2$ must be equal and opposite at $r=0$ and thus:
\begin{equation}
\rho_1(\psi)   ~=~ -\rho_2(\psi)   \,.
\label{oppdensities}
\end{equation}
This equation is the generalization of (\ref{constr1}) to a magnetube with species oscillations. On the other hand, one can also recover this equation by starting from a supertube that is not located on a $V=0$ surface \cite{Bena:2010gg}  (where the analogous condition was a linear relationship between $\rho_1(\psi)$, $\rho_2(\psi)$ and $\hat \rho(\psi)$)  and try to push this supertube onto the critical surface. Hence, a magnetube at a critical surface can be seen as an infinitely-boosted regular supertube.

Since the functions $\lambda_I$ are harmonic and fall off at infinity, (\ref{oppdensities}) implies that we must have:
\begin{equation}
\lambda_1(\psi,\vec y)   ~=~ -\lambda_2(\psi,\vec y)  \quad \Rightarrow  \quad L_1(\psi,\vec y)   ~=~ -L_2(\psi,\vec y)    \,.
\label{L1L2reln}
\end{equation}

Using (\ref{L1L2reln}) and $K^1 \equiv  K^2$, $K^3 = K^4 \equiv 0$ and the vanishing of $V$ and $L_0$  at $r=0$, the second regularity condition collapses in the relatively simple condition: 
\begin{equation}
\lim_{r  \to 0}\,  r^2\, \big[ {L_1 \, L_2 - \coeff{1}{2} \, (L_3^2 + L_4^2)}  ~-~ 2\, K^0 \, M \big] ~=~ 0  \,. 
\label{MLreln}
\end{equation}
This implies that 
\begin{equation}
2\, \alpha\, \hat \rho (\psi)     ~=~  \rho_1(\psi)\,\rho_2(\psi) - \coeff{1}{2} \, (\rho_3(\psi)^2 + \rho_4^2(\psi))  ~=~  -\big( \rho_1(\psi)^2  + \coeff{1}{2} \,(\rho_3(\psi)^2 + \rho_4^2(\psi))\big)      \,,
\label{rhohatreln}
\end{equation}
which is the analog of (\ref{constr3}).   A similar relation between $\hat \rho (\psi)$ and the other charge densities was also found in \cite{Bena:2010gg}. Note that while the integral of $\rho_I(\psi)$ vanishes, the integral of $\hat{\rho}(\psi)$ cannot (for non-trivial charge densities) and so the supertube will carry a net angular momentum. A regular supertube is then parametrized by the constant dipolar magnetic charge $\alpha$ and three independent electric charge densities, $\rho_1(\psi)$, $\rho_3(\psi)$ and $\rho_4(\psi)$. 

\subsection{Solutions with a $\psi$-independent metric}

For general charge densities, the example constructed above will have a metric that is a non-trivial function of $\psi$, although it will become $\psi$-independent asymptotically. Remarkably, however, we can choose the oscillating charge densities in such a way that  the metric is completely independent of $\psi$.  As described in the Introduction, this works in a manner rather reminiscent of $Q$-balls:  The species fluctuate but the energy-momentum tensor, and hence the metric does not.  

The basic idea is to  arrange that 
\begin{equation}
\lambda_1(\psi, \vec y)^2 + \coeff{1}{2}\,\big( \lambda_3(\psi, \vec y)^2  +  \lambda_4(\psi, \vec y)^2\big)  ~=~   \lambda(\vec y)^2   \,,
\label{lamquad}
\end{equation}
for some function $\lambda(\vec y)$. From the asymptotic properties of the $\lambda_I$, it follows that we must have
\begin{equation}
\rho_1(\psi)^2 +   \coeff{1}{2}\,\big( \rho_3(\psi)^2  +  \rho_4(\psi)^2 \big) ~=~   \beta^2   \,,
\label{rhoquad}
\end{equation}
where $\beta$ is a constant.  Then \eqref{lamquad} will make the warp factor $H$ independent of $\psi$, while \eqref{rhoquad} makes the angular momentum density determined by \eqref{rhohatreln} a constant, and hence $M$ is independent of $\psi$. Since $Z_0$ is unchanged by the addition of the supertube (as in the non-oscillating template), the full metric \eqref{sixmet1} preserves the Killing symmetry, $\frac{\partial}{\partial\psi}$ . 

It is clearly easy to choose the charge densities to satisfy \eqref{rhoquad}; this is just one constraint on the three free functions. It is a much taller order to satisfy \eqref{lamquad}, however, as this must be satisfied for all $\vec{y}$, and we only have the freedom to choose the source functions that  are functions only of $\psi$. However, the isometry of the base metric along $\psi$ means that the solution can be decomposed into Fourier modes, and if we take the source charge densities to involve a single Fourier mode, then we can take solutions $\lambda_I(\psi, \vec{y})$ that take a product form, $\lambda_I(\psi, \vec{y}) = F_n(\vec{y}) \rho_I(\psi)$, and the $\vec{y}$ dependence factors out of \eqref{lamquad} which reduces simply to \eqref{rhoquad}.\footnote{Note that this is a very special choice. Any source function can be written in terms of a superposition of Fourier modes, but if we consider more than one Fourier mode this factorization does not occur, and \eqref{lamquad} cannot be satisfied.} Thus,  we can satisfy  (\ref{lamquad}) by taking, for example, 
\begin{equation}
\lambda_1(\psi, \vec y) ~=~ \beta\, F_n(\vec y) \, \cos (n \psi)  \,, \qquad \lambda_3(\psi, \vec y) ~=~  \lambda_4(\psi, \vec y) ~=~  \beta\, F_n(\vec y) \, \sin (n \psi)  \,,
\label{Fmodes1}
\end{equation}
where $\beta$ is a constant and $F_n$ is normalized so that  $F_n\sim \frac{1}{r}$ as $r \equiv |\vec y| \to 0$.

In this way, species $1$ is locked to species $2$ and they oscillate into species $3$ and $4$.  (There is obviously a family of choices of how to distribute the oscillations amongst species $3$ and $4$ in such a way as to satisfy (\ref{rhoquad}).) We will compute the details of the functions $F_n$ in Section \ref{Sect:Modes}, but before going into the technicalities we can make some observations about the solution.

First note that (\ref{rhoquad}) and (\ref{rhohatreln}) imply that $\hat \rho (\psi)$ is constant, and  the identity (\ref{rhohatreln}) collapses precisely to that of the template solution  (\ref{constr3}).  This means that the function $M$ is exactly as in   (\ref{Mform}).  Note, in particular, that  $M$ still knows about the {\it amplitude}, $\beta$, of the species oscillation.  Furthermore, having $K^1 \equiv  K^2$, $K^3 = K^4 \equiv 0$ and $\lambda_1 =-\lambda_2$ means that the $\lambda_I$ cancel out in the function $\mu$ in (\ref{muform})  and so it is exactly the same function as it was in the template solution of Section \ref{Sect:Temp}.  Similarly, the  $\lambda_I$ cancel out in (\ref{simpomegaeqn})  and the fact that $M$ is independent of $\psi$ means that $\omega$ satisfies exactly the same equation as in the template solution and so is identical with the solution (\ref{omres}).  

The only difference between the metric of the oscillating solution and the metric of the template arises in the warp factor, $H$, given by (\ref{Hdefn}).  It follows from (\ref{ZIform}),  (\ref{sameKform}),  (\ref{newLform}), (\ref{Z34simp}) and (\ref{lamquad})  that 
\begin{equation}
H^2 ~=~ \Big[ Z_0 + \frac{\alpha}{r} \frac{K}{V} \Big]^2~-~  \lambda(\vec y)^2    \,.
\label{Hsqident}
\end{equation}
The important point is that this function is independent of $\psi$ and thus $\frac{\partial}{\partial \psi}$ is still a Killing vector of the metric even though the Maxwell fields and scalars are $\psi$-dependent.  Because of (\ref{rhoquad}) the function $H$ also has exactly the same behavior as $r \to 0$ as its counterpart (\ref{Htemp}) in the template solution.  Thus the local regularity of the metric is guaranteed by that of the template solution. 

The fact that the metric is unchanged apart from this change in the warp factor $H$ also implies that leading-order asymptotics of the metric are also the same as for the non-oscillating solution. Indeed, the asymptotics is independent of the assumption that the full metric is $\psi$-independent. Any $\psi$-dependence in $H$ is subleading at large distances, and thus so long as \eqref{rhoquad} is satisfied so that the angular momentum is $\psi$-independent, the leading-order asymptotics will be the same as in the non-oscillating solution. Thus, for all solutions satisfying \eqref{rhoquad}, the overall amplitude of the oscillating charge densities is bounded by \eqref{bound1}.

Furthermore, the $\lambda(\vec{y})^2$ in \eqref{Hsqident} will not contribute to the asymptotic stress tensor: as we we will see in the next subsection, the functions $F_n(\vec{y})$ represent higher multipole moments and therefore fall off faster than $1/r$, so $\lambda(\vec{y})^2$ in \eqref{Hsqident} falls off faster than $1/r^2$. Thus this term falls off too fast to contribute to the sub-leading part of the metric that gives the asymptotic charges. For the gauge fields, there was no contribution to charges in the non-oscillating case due to the absence of a logarithmic term in \eqref{deltaA}; the faster fall-off in the oscillating solution does not change this. Thus, the asymptotics of solutions with species oscillation where the angular momentum density is a constant is very similar to the non-oscillating case. The only notable difference in the asymptotics is that the suppression of  $F_n(\vec{y})$ implies that the Wilson line noted in \eqref{deltaA} is absent in the oscillating solution.

\section{The oscillating modes}
\label{Sect:Modes}

\subsection{The Green's function}

To obtain the explicit details of our oscillating solution we need to solve (\ref{Lharm}) on the base defined by (\ref{GHmetric}) with (\ref{Vform}).  Fortunately the relevant Green's function was computed in \cite{Bena:2010gg}.  The expression for this function was very complicated but here there are significant simplifications because we have put the source point, $(\psi',\vec y)$,  at $\vec y =0$. 

Define the following combinations of coordinates:
\begin{equation}
u  ~\equiv~  \cosh \xi    \,,  \qquad y   ~\equiv~     \sin \theta   \,, \qquad
w  ~\equiv~ e^{{i \over 2 q} (\psi - \psi') -  i (\phi - \phi')}   \,,
 \label{params1}
 \end{equation}
and  set 
\begin{align}
w_\pm   &~\equiv~  \frac{1}{2\, u^2 \,y}\,\big[\,(2\, u^2 -1 + y^2)  ~\pm~ \sqrt{1 - y^2} \,\sqrt{(2\, u^2 -1)^2 - y^2}\,  \big] \,,  \label{params2} \\
x    &~\equiv~ \coeff{1}{2}\, (w_+ + w_-) ~=~ \frac{(2\, u^2 -1 + y^2) }{2\, u^2 \,y} \,,
 \label{params3}
 \end{align}
Observe that $w_+ w_- =1$ and hence $x \ge 1$ with equality if and only if $|w_\pm| =1$.  We will adopt the convention of taking the positive square roots in (\ref{params2}) so that $|w_-| \le 1 \le |w_+|$.

It is also convenient to recall that one can map back to the polar variables $r_\pm$ and $r$ by using (\ref{coordsa})  and (\ref{rpmform}):
\begin{equation}
u^2   ~=~  \frac{1}{4\,a} \, (r_+ + r_- + 2a)     \,,  \qquad  y  ~=~  \frac{1}{2\,a} \, \sqrt{(r_+ + r_- -2 r)(r_+ + r_- + 2 r)}     \,.
 \label{invparams1}
 \end{equation}
Thus the end result can be expressed as a rational function of these variables.  In particular, it is  useful to note that 
\begin{equation}
\sqrt{(2\, u^2 -1)^2 - y^2}  ~=~  \frac{r}{a}    \,.
 \label{invparams2}
 \end{equation}

The  Green's function can then be written as \cite{Bena:2010gg} 
\begin{align}
\widehat G ~=~ & - \frac{1}{a\, u \,y}\, {\rm Re}\bigg[\,  \frac{ w (1- y\, w) }{(w- w_+) (w - w_-)} \, \frac{ 1 }{\sqrt{u^2 - y \, w}}\, \bigg]\,. \label{Gnice1} \\
~=~ & - \frac{1}{2\, a\, u \,y}\,\frac{ w }{(w- w_+) (w - w_-)} \,  \bigg[\,  \frac{(1- y\, w) }{\sqrt{u^2 - y \, w}}~+~ \frac{  (w- y) }{\sqrt{u^2 - y \, w^{-1}}}\, \bigg]\,,\label{Gnice2} 
 \end{align}
where we have used the fact that $w^* = w^{-1}$.

There are several important things to note:  (i) $r=0$ corresponds to $u=1$ and $y =1$ and we are going to work in the region $r>0$ in which $y^{-1} u^2 > 1$.  (ii) The branch points of $\sqrt{u^2 - y \, w}$ are  at infinity and at $w = y^{-1} u^2 > 1$ and since $|w|=1$, the branch cuts can be arranged to be safely outside the unit circle.  (iii) The fact that $w_+ w_- =1$ means that one of the poles at $w_\pm$ is outside the unit circle while the other is inside.  We are choosing signs of square roots so that  $|w_-| \le 1 \le |w_+|$.  (iv) The poles at $w_\pm$ coincide at $1$ if and only if $y=1$.  This is the critical surface, and a careful examination of (\ref{Gnice2}) shows that $\widehat G =0$ when $y=1$ and is indeed well-behaved across this surface. 

\subsection{The Fourier modes}

\subsubsection{The explicit modes}
\label{Sect:Explicit}

One can obtain the Fourier modes from this Green's function by integrating against $e^{-i n (\frac{1}{2q}\psi' -\phi')} = w^n e^{-i n (\frac{1}{2q}\psi -\phi)}$.  Dropping the phase factors of $e^{-i n (\frac{1}{2q}\psi -\phi)}$, the functions of interest are 
\begin{equation}
F_n ~=~  \frac{1}{2\pi i} \, \oint_{|w| =1} \, \frac{d w}{w} \, w^n \big(  F(w) ~+~ F(w^{-1})\big)   \,, \label{contour1} 
 \end{equation}
where the contour is taken counterclockwise around the unit circle and
\begin{equation}
F(w)  ~\equiv~  -\frac{1}{2\, a\, u \,y}\, \bigg[\,  \frac{ w (1- y\, w) }{(w- w_+) (w - w_-)} \, \frac{ 1 }{\sqrt{u^2 - y \, w}}\, \bigg]\,.  \label{Fdefn} 
 \end{equation}
Since $F(w^{-1})$ has a branch cut inside the unit circle, we make the inversion, $w \to w^{-1}$ in this integral (remembering that the inversion also inverts the orientation of the contour) and write it as
\begin{equation}
F_n ~=~  \frac{1}{2\pi i} \, \oint_{|w| =1} \, \frac{d w}{w} \,    ( w^n ~+~  w^{-n}) \, F(w)     \,. \label{contour2} 
 \end{equation}
There are now only contributions from poles at $w=0$ and at $w=w_-$.  We can also assume, without loss of generality, that $n \ge 0$.

An identity that is useful in evaluating residues at $w=w_-$ is 
\begin{equation}
\sqrt{u^2 - y \, w_- }~=~    \frac{u\,  (1 - y\, w_-)  }{\sqrt{1 - y^2}}   \,. \label{sqrtsimp} 
 \end{equation}
where we are again taking the positive square roots and have used  $|w_-| \le 1$.

The residue at $w_-$ is now easy to compute:
\begin{equation}
F_n^{(w_-)} ~=~   \frac{1}{2\,r} \, (w_+^n +w_-^n)    \,. \label{Fnwminus1} 
 \end{equation}
Note that for $n=0$ this simply gives $r^{-1}$ and this has been used to set the normalization of $F_{n}$ in general.  For general $n$ we can write this in terms of Chebyshev polynomials:
\begin{equation}
F_n^{(w_-)}   ~=~   \frac{T_n(x)}{r}  \,. \label{Fnwminus2} \,,
 \end{equation}
where $T_n$ is the  $n^{\rm th}$ Chebyshev polynomial of the first kind and $x$ is given by  (\ref{params3}).  

To evaluate the residue at $w=0$, we can use the expansion
\begin{equation}
\frac{1}{w^2 - 2\, x \,w +1} ~=~  \sum_{p=0}^\infty \, U_p (x) \, w^p \,, \label{ChebyGen} 
 \end{equation}
where $U_p (x)$ are Chebyshev polynomials of the second kind.  This series converges for $|w| < |w_-|$ and hence near $w=0$.  The residue at $w=0$ can then be written as
\begin{equation}
F_n^{(0)} ~=~ -\frac{1}{2\, a\, u^2 \,y}\, \sum_{k=0}^{n-1}  \, \frac{1}{k!} \, U_{n-k-1} (x)  \, \bigg[ \frac{d^k}{d^k w}\bigg( \frac{  (1- y\, w) }{\sqrt{1 - \frac{y}{u^2} \, w}} \bigg) \bigg] \bigg|_{w=0} \,. \label{Fn0} 
 \end{equation}
Thus the solutions to (\ref{harmlam}) can be written as linear combinations of the modes:
\begin{equation}
F_n  \, e^{-i n (\frac{1}{2q}\psi -\phi)} ~=~ \big( F_n^{(0)}  ~+~F_n^{(w_-)}\big)\,  e^{-i n (\frac{1}{2q}\psi -\phi)}  \,. \label{Fnmodes} 
 \end{equation}
 with $F_n^{(0)}$  and $F_n^{(w_-)}$ given by (\ref{Fn0}) and (\ref{Fnwminus2}).  In particular, $F_n^{(w_-)}$ gives rise to the singular source at $r=0$ and the remaining parts are essential corrections as we now discuss.
 
\subsubsection{On-axis limit}
\label{Sect:Axis}

This description of the modes appears to be rather singular at $y=0$, particularly given the form of $x$ in (\ref{params3}).  This limit corresponds to $\theta = 0, \pi$ and represents the axis through all the source points. 

It is easy to see that this apparent singularity is an artifact of how we assembled the modes.  Indeed, one can easily take the limit of $y \to 0$ in (\ref{Fdefn}) and define
\begin{equation}
\hat F(w)  ~=~ \lim_{y \to 0} \,F(w) ~=~   \frac{1}{2 \,a\, (2u^2 -1)} \,. \label{hatFdefn} 
 \end{equation}
and hence 
\begin{equation}
F_n ~\to~   \frac{1}{2 \,a\, (2u^2 -1)}\,\frac{1}{2\pi i} \,  \oint_{|w| =1} \, \frac{d w}{w} \,    ( w^n ~+~  w^{-n})  ~=~  \frac{1}{a\, (2u^2 -1)}\, \delta_{n,0}   \,. \label{contour3} 
 \end{equation}
which is manifestly smooth and, in fact, only non-zero for $n=0$.  

One can therefore take the view that $F_n^{(w_-)}$ has the role of generating the proper singular source at $r=0$ and then  $F_n^{(0)}$ has the role of fixing all the other unphysical singularities in $F_n^{(w_-)}$.

\subsubsection{Examples}
\label{Sect:Examples}

As we have already seen,
\begin{equation}
F_0 ~=~ \frac{1}{a\, \sqrt{(2\, u^2 -1)^2 - y^2}} ~=~   \frac{1}{r}    \,, \label{F0works} 
 \end{equation}
and this was used to set the normalization of the modes.

The first few Chebyshev polynomials are 
\begin{equation}
T_0 ~=~  1\,, \quad T_1 ~=~  x \,, \quad T _2 ~=~  2 x^2 -1 \,; \qquad U_0 ~=~  1\,, \quad U_1 ~=~  2x \,, \quad U_2 ~=~  4 x^2 - 1  \,. \label{Chebys} 
 \end{equation}
Thus 
\begin{align}
F_1 &~=~ \frac{T_1(x)}{r} -  \frac{1}{2 \,a\, u^2\, y } \, U_0(x) \, \\ 
&~=~    \frac{1}{r} \, \Big(\frac{r_+ + r_- + 2r + 2a }{r_+ + r_- + 2a}\Big) \, \sqrt{\frac{r_+ + r_- - 2r}{r_+ + r_- + 2r}}      \,, \label{F1form} 
 \end{align}
where we have used  (\ref{invparams1}) and  (\ref{invparams2}).
Similarly, one has
\begin{align}
F_2 &~=~ \frac{2 x^2 -1}{r} -  \frac{1}{2 \,a\, u^4\, y } \, \big(2 \, u^2\, x  - \coeff{1}{2}\, (2 \, u^2 -1)\big) \\ 
&~=~   \Big(\frac{r_+ + r_- - 2r}{r_+ + r_- + 2r}\Big) \, \bigg(\,  \frac{1}{r}  + \frac{4(2r -a) }{(r_+ + r_- + 2a)^2} +  \frac{6}{(r_+ + r_- + 2a)}   \, \bigg)  \,.\label{F2form} 
 \end{align}
Note that both $F_1$ and $F_2$ vanish on the axis, where $ r_+ + r_- = 2 r$.  Moreover they both have a canonically normalized source at $r=0$:
\begin{equation}
\lim_{r \to 0} r\, F_1 ~=~ \lim_{r \to 0} r\, F_2 ~=~ 1 \,.\label{sourcenorm} 
 \end{equation}
At infinity one should note that  (\ref{invparams1}) implies that as $r \to \infty$,
\begin{equation}
r_+ + r_- - 2r  ~\sim~\frac{a^2 \, \sin^2 \theta}{r} \,, \label{asympr} 
 \end{equation}
and thus
\begin{equation}
F_1 ~\sim~ \frac{a \, \sin \theta}{r^2} \,, \qquad F_2 ~\sim~ \frac{3\, a^2 \, \sin^2 \theta}{2\, r^3}   \,.\label{asympF12} 
 \end{equation}
%


\subsection{Asymptotics of the modes at infinity}

From the explicit examples above one sees that the modes $F_n \sim 1/r^{n+1}$ as would be expected from charge multipoles. We can easily prove this form in general by recalling that the Green's function \eqref{Gnice2} was obtained in \cite{Bena:2010gg} by taking the Green's function on the $AdS_3 \times S^2$ space \eqref{AdS3S2} and Kaluza-Klein reducing to obtain the Green's function on the ambi-polar base. The asymptotic behaviour of the Fourier modes of the Green's function on the $AdS_3 \times S^2$ space is then easily obtained by a conventional spherical harmonic analysis on the $S^2$. 

From the perspective of $AdS_3 \times S^2$, since $\varphi_2 = \phi - \frac{1}{2q} \psi + 2\tau$, our Fourier expansion $G = \sum_n F_n(u,y) w^n$ corresponds to an expansion 
\begin{equation}
G = \sum_n F_n(\xi, \theta) e^{-in(\varphi_2 - \varphi_2')+2in (\tau-\tau')}.
\label{GfnF}
\end{equation}
The Green's function can be decomposed in terms of the spherical harmonics $Y_{lm}(\theta,\varphi_2)$ on the $S^2$; in this decomposition, $F_n$ will only involve spherical harmonics with $l \geq n$. Doing a Kaluza-Klein reduction on the $S^2$, the massless wave equation on $AdS_3 \times S^2$ becomes a massive wave equation on $AdS_3$ for each spherical harmonic mode with a mass $m^2 = l(l+1)/R_2^2= 4l(l+1)/R_1^2$. Thus, $F_n = \sum_{l > n} F_{ln}$ where at large $\xi$
\begin{equation}
\sinh 2\xi^{-1} \partial_\xi \sinh 2\xi \partial_\xi F_{ln} \approx -4l (l+1) F_{ln}, 
\end{equation}
so $F_{ln} \sim e^{-2(l+1) \xi}$, that is at most $F_n \sim 1/r^{n+1}$.  From the perspective of the $AdS_3 \times S^2$ asymptotics, the species oscillation of the supertube is introducing a variation of the fields along the $\varphi_2$ direction in the  $S^2$, so it corresponds to exciting higher KK harmonics, which produce vevs for higher-dimension operators in the dual CFT rather than exciting local charge densities of the conserved currents.

\section{Conclusions}
\label{Sect:Conclusions}

\subsection{Comments on supersymmetric species oscillation}
\label{BPSConc}

Our aim has been to show that the object underlying the species oscillation idea introduced in \cite{Mathur:2013nja} is a supersymmetric magnetube: a combination of (M5 and P) magnetic charges as well as several oscillating (M2) electric charges, that carries the same supersymmetries as M5 branes and momentum regardless of the oscillations of its electric charges. Trying to bend this object into a ring in $\IR^4$ will break these supersymmetries, and will result in neutral configurations that have been proposed in \cite{Mathur:2013nja} to describe microstate geometries of neutral Schwarzschild black holes. 

We have also succeeded in realizing the species oscillation idea in a supersymmetric context, by embedding the magnetube into supersymmetric solutions. In doing this we have uncovered a remarkable unification of the ``timelike'' and ``null'' types of supersymmetries directly in the framework of five-dimensional supergravity; the ``timelike'' supersymmetry on an ambi-polar background becomes locally null on the surface $V=0$. Hence an object that would be invariant only under the magnetic-type ``null'' supersymmetries, like the magnetube, can be placed on this surface while preserving the global electric-type ``timelike'' supersymmetry. This enables us to construct solutions with species oscillation which preserve supersymmetry globally. 

We have carefully analyzed the smoothness conditions for magnetubes on the $V=0$ surface, initially for a magnetube without species oscillation and then for a magnetube with M2 electric charge densities that vary around the tube in such a way that the net M2 charges vanish. We found that this magnetube can be obtained by moving a supertube onto the $V=0$ surface (which corresponds to giving it an infinite boost) and that, in the $V \rightarrow 0$ limit, the smoothness conditions of the supertube are enough to ensure the smoothness of the magnetube. We constructed a simple example of a solution with species oscillation and gave an explicit description of its structure. In this  example, the geometry is independent of the $\psi$ coordinate; the gauge fields are oscillating but their combination in the stress tensor is $\psi$-independent, as in the exact near-tube solution discussed in \cite{Mathur:2013nja}. 

The simple example we have considered is asymptotically $AdS_3 \times S^2$, and it would be interesting to understand  the interpretation of the solution with species oscillation from the point of view of the dual CFT.  A key aspect is that when we introduce the $\psi$-dependent source, the non-diagonality of the metric in the angular coordinates makes the solution a function of $\frac{\psi}{2q} - \phi$, as can be seen in the Green's function \eqref{Gnice1}. This combination becomes an angle in the $S^2$ factor in the asymptotic solution \eqref{AdS3S2}. Thus, from the point of view of the $AdS_3 \times S^2$ background, the species oscillation is not introducing dependence on the angular coordinate in the $AdS_3$ (shifting $\varphi_1$ at fixed $\varphi_2$ remains a symmetry of the solution) but on one of the angular coordinates on the $S^2$. Thus, understanding the interpretation of our solution in the dual CFT will involve a Kaluza-Klein decomposition along the lines of \cite{Skenderis:2006uy}. 

It will obviously be interesting to further exploit this new freedom in constructing supersymmetric smooth microstate geometries. It gives us a new possibility to introduce dipole charges. Examples of solutions with varying charge densities were previously constructed in \cite{Bena:2010gg}, but we can now construct solutions with zero net charge. Another interesting direction for further development is to consider species oscillation on supersymmetric black strings and black rings.

\subsection{Non-BPS, asymptotically flat solutions}
\label{nonBPS}

Following the direction of \cite{Mathur:2013nja}, probably the most interesting question to ask is if one can deform our supersymmetric solutions to construct non-supersymmetric asymptotically-flat solutions, as a step towards constructing microstates for black holes like Schwarzschild. 

The solution we constructed above is asymptotically $AdS_3 \times S^2$, but one should be able to construct similar supersymmetric solutions that are asymptotically flat in five dimensions by considering a supertube/magnetube at the $V=0$ surface in a more general Gibbons-Hawking base that gives rise to an asymptotically flat background. We have also seen that the oscillating charge distributions do indeed make no net contribution to the charges measured from infinity. On the other hand, the size of the oscillation is still bounded by the background magnetic fluxes and hence by the asymptotic charges.

It is remarkable that one can have stationary asymptotically flat solutions; one might physically expect that a varying charge density on a rotating object would give rise to time-dependent multipole moments, which would lead to electromagnetic radiation. However, the resolution is a familiar effect in supertubes; the supertube is rotating, but the charge on the supertube corresponds to a charge density wave traveling around the tube in the opposite direction, so that we get a standing wave. As a result, the multipole moments of our supersymmetric solutions, whose oscillating parts may be viewed as superubes/magnetubes, will all be time-independent. 

On the other hand, this cancellation of the time-dependence coming from having a standing wave is a fine-tuned phenomenon. If we excite the supertube/magnetube by adding some energy, we will need to make it rotate faster to maintain the stabilization by angular momentum, and  it is possible that the counter-rotating charge density wave will then give rise to time-dependent multipole moments. If true, this would a real obstacle to finding stationary non-supersymmetric solutions. Note also that the non-stationarity is associated with the emission of electromagnetic radiation, so the natural timescale for the decay is likely to be much faster than the gravitational timescales associated with a black hole.

The solution discussed in \cite{Mathur:2013nja} is vulnerable to a similar problem. In the exact near-tube solutions the authors introduce dependence on a null coordinate $t - a\phi$, corresponding to a charge density that is a function of $t-a\phi$. In the near-tube solution $\phi$ is a coordinate along a straight black string, and this is related to a solution with charge density depending just on the position along the string by an (infinite) boost. But in the full asymptotically-flat solution this corresponds to a truly time-dependent charge distribution, and we expect that this will lead to electromagnetic radiation in the sub-leading corrections to the solution, introducing time-dependence in the metric on a timescale set by the electromagnetic radiation reaction.  

Thus we believe that species oscillation has opened up an even larger, new moduli space of BPS microstate geometries but remain uncertain whether this idea can be adapted to yield stationary non-BPS configurations.
 
\bigskip
\leftline{\bf Acknowledgements}
\smallskip
We would like to thank Samir Mathur and David Turton for extensive discussions and for providing us with an advance copy of \cite{Mathur:2013nja}. We would also like to thank Stefano Giusto, Rodolfo Russo and  Masaki Shigemori for useful discussions.  NPW is grateful to the IPhT, CEA-Saclay and to the Institut des Hautes Etudes Scientifiques (IHES), Bures-sur-Yvette, for hospitality while this work was done. NPW would also like to thank the Simons Foundation for their support through a Simons Fellowship in Theoretical Physics. SFR would like to thank the UPMC for hospitality  while this work was done. NPW and SFR are also grateful to CERN for hospitality  while this work was initiated, and we are all grateful to the Centro de Ciencias de Benasque for hospitality at the ``Gravity- New perspectives from strings and higher dimensions'' workshop. The work of IB was supported in part by the ANR grant 08-JCJC-0001-0, and by the ERC Starting Independent Researcher Grant 240210-String-QCD-BH and by a grant from the Foundational Questions Institute (FQXi) Fund, a donor advised fund of the Silicon Valley Community Foundation on the basis of proposal FQXi-RFP3-1321 to the Foundational Questions Institute. This grant was administered by Theiss Research. The work of NPW  was supported in part by the DOE grant DE-FG03-84ER-40168. SFR was supported in part by STFC and the Institut Lagrange de Paris.




\end{document}